\documentclass[11pt]{article}

\usepackage{amssymb}      
\usepackage{tabularx}     
\usepackage{xcolor}       
\usepackage{graphicx}  
\usepackage{makeidx}
\usepackage{times}
\usepackage{amssymb}
\usepackage{amsmath}
\numberwithin{equation}{section}
\usepackage{setspace}
\usepackage{color}
\usepackage{graphicx}
\usepackage{rotating}
\usepackage{pdflscape}
\usepackage{epstopdf}
\usepackage[round,authoryear,comma]{natbib}
\usepackage{natbib,hyperref}
\usepackage[margin=1in]{geometry}

\usepackage{booktabs}
\usepackage{pgfplots}
\usepackage{subcaption}
\usepackage{multirow}
\usepackage{array}
\usepackage{placeins}
\usepackage{actuarialangle}
\hypersetup{
	colorlinks   = true,
	urlcolor     = blue, 
	linkcolor    = blue,
	citecolor   = blue 
}
\providecommand{\U}[1]{\protect\rule{.1in}{.1in}}
\parskip=\medskipamount

\newcolumntype{C}[1]{>{\centering\arraybackslash}m{#1}}
\begin{document}

\title{Liquidity Shocks, Homeownership, and Income Inequality: Impact of Early Pension Withdrawals and Reduced Deposit}
\date{Version: \today}
	\author{Hamza Hanbali \\ \small Department of Economics, Centre for Actuarial Studies\\ \small University of Melbourne\\  \texttt{{\small hamza.hanbali@unimelb.edu.au}} \and Gaurav Khemka \\ \small Research School of Finance, Actuarial Studies \& Statistics \\ \small Australian National University\\ \texttt{\small gaurav.khemka@anu.edu.au} \and Himasha Warnakulasooriya\\ \small Department of Econometrics and Business Statistics \\ \small Monash University \\ \texttt{\small himasha.warnakulasooriya@monash.edu}}
\maketitle
\bigskip\bigskip
\begin{center}
	\hrule
\end{center}
\begin{abstract}
\noindent The paper analyzes two government policies affecting housing demand: early withdrawal from pension savings (EW), and reduction of loan deposit (RD). A model incorporating demand feedback on housing prices using Australian data shows both policies raise prices in the short run. RD delays or prevents access for low-income households, particularly in supply-constrained markets. EW improves accessibility across groups and is most efficient when full withdrawal is permitted, but can reduce retirement security if pension grows faster than property prices. The results also indicate that unequal outcomes stem not from price surges themselves but from pre-existing market disparities.
\end{abstract}
\begin{center}
	\hrule
\end{center}

\newpage
\section{Introduction}
%
%

%
%
%
%
%

\label{Introduction}
Homeownership is elusive for many households despite being a central pillar of wealth accumulation and financial security \citep{2022AtalayEdwards,2023Sodini}. Governments worldwide have implemented various subsidies and tax incentives to facilitate access to homeownership. Some of the most popular programs include tax credits or interest deductions, which are known for driving up house prices, and benefiting sellers rather than improving homeownership \citep{2006Vigdor,2013Bourassa,2013Fetter,2014HilberTurner,2018SommerSullivan,2020BergerTurnerZwick,2023Krolage}. This price inflation arises because demand-side interventions increase purchasing power without necessarily increasing supply, particularly in housing markets characterized by inelastic supply \citep{2015Favara,2024Carozzi}. What remains underexplored is the distributional impact of these policies: Do liquidity shocks that facilitate home purchases exacerbate inequality rather than alleviate it? 

The present paper considers two housing policies designed to reduce the upfront cost of homeownership. The first is a reduction in the minimum deposit requirement, supported by a government guarantee (henceforth \textit{reduced deposit}, or RD). The second combines a reduced deposit threshold with early withdrawal from pension savings to cover the remaining amount (henceforth \textit{early withdrawal}, or EW). Both policies can be understood as liquidity shocks, though of a different nature from traditional government- or tax-funded subsidies. Importantly, they do not directly increase households’ total wealth, but instead relax liquidity constraints by either reducing the required cash contribution (RD) or unlocking otherwise illiquid retirement savings (EW).

Specifically, the RD policy lowers the immediate liquidity requirement for homeownership by reducing the deposit from the standard 20\% (a common benchmark in countries such as Australia, Canada, the United Kingdom, and the United States) to just 5\%. While low-deposit mortgages (5\% or less) are available in these countries, they typically come at a cost, such as mortgage insurance premiums or higher interest rates. In contrast, the RD policy considered in this paper assumes a government guarantee and imposes no additional borrowing costs on the household.

On the other hand, the EW policy grants individuals access to their own tax-advantaged retirement savings. These savings are typically locked until retirement, but since homeownership itself reduces housing costs in old age, this policy could accelerate that benefit by enabling earlier access to secure housing now rather than later. Early access to pension savings for housing is only feasible within pension systems that assign a specific account balance to each individual, typically Defined Contribution (DC) systems. Unlike Pay-As-You-Go (PAYG) schemes, where current contributions fund current retirees, DC systems accumulate savings in individual accounts that can be accessed under certain conditions. Some countries with mandatory or large-scale DC systems allow housing-related pension withdrawals. This includes Kazakhstan and Singapore, as well as Switzerland, where a legally Defined Benefit (DB) system functions in practice like a DC scheme by allocating individual balances and permitting early withdrawals for homeownership. Other countries, including Canada, New Zealand, and  the United States, have shifted towards DC arrangements but have not fully mandated them at the national level. These three countries also allow pension withdrawal under some restrictions. As more countries transition towards hybrid or mandatory DC models, the debate over whether pension savings should be accessible for homeownership is becoming increasingly relevant \citep{2024Mercer}. 

The present paper develops a model of a cohort of renters, segmented by income percentiles, who accumulate savings and pension balances over time in order to purchase a home. A household purchases a property once it can afford the required minimum, which includes the deposit, property transfer taxes, and an additional financial buffer. Affordability is evaluated under each policy scenario (EW or RD) and compared to a benchmark scenario in which households rely solely on savings with a standard deposit requirement. The policies are analyzed under two housing market structures: an \textit{equal-affordability} market, where all income groups can access properties priced at the same income multiple, and a \textit{supply-constrained} market, where higher-income households have proportionally greater access to affordable properties than lower-income groups.

The model incorporates an econometric framework for key economic variables (house prices, rent, wages, inflation, borrowing rates, savings rates, and pension returns) which interacts with the lifecycle model through a demand variable. The present study does not aim to assess whether housing policies affect demand; instead, consistent with existing research on demand-side interventions, it assumes that the policies affect demand and, consequently, house prices. The effect of demand is estimated using historical data. Based on this estimated relationship, demand patterns from the simulated population at each period feed into the econometric model, which then adjusts housing prices, creating a feedback loop between household behavior in the lifecycle model and market conditions in the econometric model. The model is calibrated using Australian data and tracks a cohort of 25-year-old employed renters with no savings. The model does not allow inheritance or external financial support and neither accounts for multiple cohorts nor the advantage of those with higher existing savings at the time of implementation. As a result, the model isolates the effect of the policy on a cohort with existing income disparities.

The policy’s effects are evaluated across three dimensions: (i) household financial outcomes, including the likelihood of purchase, the timing of home purchases, and retirement financial security (i.e. retirement income net of housing); (ii) distributional impacts, measured by changes in homeownership access inequality and post-retirement financial security using Gini coefficients; and (iii) fiscal implications for governments, assessed through the net present value of tax revenues and government expenditures.

The focus on Australia in this paper is motivated by two main factors. First, Australia’s retirement system, known as superannuation, is one of the largest and most mature DC systems in the world \citep{2024Eslake, 2024Mercer}. Second, both housing policies analyzed here were prominent in the lead-up to the 2025 federal election. An EW policy was first introduced by the Liberal-National Coalition in 2022 and re-emerged during the 2025 campaign alongside proposals for mortgage interest deductibility. Supporters argue that EW promotes financial autonomy, supports homeownership, and reduces housing costs in retirement. Critics argue that it contributes to housing price inflation and compromises long-term retirement adequacy by diverting savings toward current homeowners and developers \citep{2024Eslake,2025SMC}. The RD policy was proposed by the Labor Party, which won the 2025 election. While both policies aim to lower the effective deposit barrier, EW still requires a top-up from accumulated superannuation savings, making it more restrictive. As a result, RD is expected to exert an upward pressure on housing prices earlier than EW, given its more immediate reduction in liquidity constraints.

The paper finds that, as expected, both policies raise property prices, with a higher peak under EW, but earlier surge under RD. Under the benchmark EW design, modelled on Australian settings (allowing up to 40\% withdrawal with a 5\% savings contribution), results indicate a modest reduction in the average age of purchase and an increase in purchase probability. By contrast, RD lowers purchase probabilities for lower-income households and widens disparities in entry timing, with particularly adverse effects under supply constraints. These findings remain robust when demand sensitivity in the price equation is doubled or halved. Boundary EW designs (i.e. full withdrawal or no required savings contribution) substantially improve housing accessibility. Restricting access to below-median or bottom-quartile incomes does not materially change outcomes for the unrestricted group. Retirement outcomes depend critically on pension returns. When, as observed historically, pension returns exceed house price growth, retirement security declines under EW. When the two rates are equal, both policies improve retirement outcomes. Overall, RD worsens accessibility by reinforcing inequality and pricing out lower-income households, whereas EW enhances accessibility without disadvantaging low-income groups and is most effective when full withdrawal is permitted. However, allowing full withdrawal may jeopardise retirement security if pension returns substantially exceed property growth.

This paper contributes to the existing studies on EW policies in DC pension systems. The closest related analysis is the report by \cite{2021MckellInstitute}, which warns that allowing superannuation withdrawals for housing in Australia would primarily inflate property prices without meaningfully increasing homeownership rates. The findings in this paper do not support that conclusion, and instead show that despite the feedback effect between property price and demand, EW can improves housing accessibility without disadvantaging low-income groups. Qualitative discussions in \cite{2024Eslake} and \cite{2025SMC}, including comparisons with New Zealand’s experience, raise concerns that such policies may exacerbate inequality due to unequal pension balances. The present paper finds that the negative effect on retirement financial security depends on future pension results, and that the Coalition's EW design does not lead to substantial inequality shifts.

This paper adds to the extensive literature examining the impact of liquidity shocks and housing subsidies on homeownership and property prices. Prior studies have shown that demand-side interventions, such as mortgage credit expansions \citep{2006Vigdor,2015Favara}, government-backed homebuyer subsidies \citep{2020BergerTurnerZwick,2023Krolage,2024Carozzi}, and tax incentives like mortgage interest deductions \citep{2013Bourassa,2014HilberTurner,2015BinnerDay}, tend to increase housing prices without significantly improving homeownership rates, especially in supply-constrained markets. The present paper contributes to this discussion by showing that EW belongs to the class of policies that inflate prices while significantly improving access, whereas RD policies not only increase prices but also reduce the likelihood and delay the timing of homeownership for low-income households. 

While many existing studies focus primarily on house price response, this paper goes further by quantifying the distributional consequences that arise through feedback loops between household demand and macroeconomic variables. Among the closest studies in the literature, \cite{2014HilberTurner} show that mortgage interest deductions (MID) disproportionately benefit higher-income households and, in supply-constrained markets, are fully capitalized into house prices, reducing accessibility for lower-income groups. Similarly, \cite{2018SommerSullivan}, which is closest to the present paper in their methodology, demonstrate that removing the MID leads to lower prices and higher homeownership rates, particularly for lower-income households. 

The present paper extends \citet{2014HilberTurner} and \citet{2018SommerSullivan} in several respects. First, it studies policies that differ meaningfully from the MID by operating before purchase and relaxing upfront liquidity constraints, and it examines cases where access is restricted by income. The results show that RD's distributional effects resemble those of the MID (i.e. reinforcing inequality under supply constraints), whereas EW can be beneficial. This distinction refines the understanding of how price-inflating housing policies affect welfare, by showing that rising prices are not uniformly harmful. Specifically, although both RD and EW raise prices in the short run, RD delays or prevents access for lower-income households, and EW advances purchase timing across groups with limited downside for inequality. More importantly, the findings show that market structure is in large part responsible for the incidence on inequality, not the price surge itself. Equal-affordability versus supply-constrained environments produce different shifts, indicating that pre-existing market disparities drive unequal outcomes. This conclusion is reinforced by robustness to doubling or halving the demand parameter in the price equation.

The remainder of the paper is structured as follows. Section \ref{Section-model} presents the general life-cycle model. Without relying on country-specific details, it focuses on the general mechanics linking household decisions to economic inputs and outcomes. Section \ref{Section-Sim} introduces the assumptions applied to input variables, covering taxes, the pension system, and macroeconomic dynamics, as well as the policy scenarios. These assumptions are grounded in the Australian institutional context, and the housing policies reflect the Liberal-Nationals Coalition's proposal for early superannuation withdrawal, and the Labor Party's proposal for a reduced deposit. Section \ref{Sec:Results} presents the results under the baseline modeling assumptions, as well as robustness tests under different assumptions on the effect of demand and pension returns. Section \ref{Sec:Alternatives} explores the case where the policies' access is restricted by income, as well as boundary cases for EW where households can withdraw all the pension balance, or where no savings contribution is required from the household. Section \ref{Sec:Conclusion} concludes.

\section{General life-cycle model}\label{Section-model}

This section presents a life-cycle model designed to be general and applicable across different countries. Country-specific assumptions are introduced in Section \ref{Section-Sim}, where the model is calibrated to the Australian case.

The structure of this section is as follows. Section \ref{Sec:Model1} defines preliminary time and population variables. Section \ref{Sec:Model2} introduces the core equations governing the evolution of savings and pension account balances, along with other related variables. Section \ref{Sec:Model3} defines the purchase time and provides the corresponding withdrawal rules for the savings and pension accounts.

\subsection{Preliminaries}\label{Sec:Model1}
Time is modeled as discrete and denoted by $t$, where period $t$ represents the interval $[t, t+1)$. Without loss of generality, all variables are assumed to remain constant within each period. Transactions occur at the beginning of each period. Households exit the population either due to death or default, both of which are assumed to occur at the end of the period. 

Retirement time is denoted by $T_r$, and property purchases are funded through loans with a fixed term  $T_l$. Time of death is denoted by $T_{\sigma}$. Default time is denoted by $T_d$, and is formally introduced in Section \ref{Sec:Model2}. Purchase time, denoted by $T_p$, is an outcome of the model defined in Subsection \ref{Sec:Model3}.

The population consists of $N(0)$ households at the start of the analysis. The survival status of household $i=1,...,N(0)$ at time $t$ is denoted by $\sigma^{(i)}(t)$, and is equal to 1 if at least one member is still alive, and to 0 otherwise. The default status of household $i$ at time $t$ is denoted by $d^{(i)}(t)$, and is equal to 1 if the household has defaulted before time $t$, and to 0 otherwise. Default is assumed to occur when household's savings reach zero, where savings include equity (the paid proportion of the home loan), but exclude pension fund balance. The number of households in the population at time $t$ is denoted by $N(t)$:
$$N(t)=\sum_{i=1}^{N(0)}\sigma^{(i)}(t)\times (1-d^{(i)}(t)).$$

Households are planning for homeownership by accumulating a minimum deposit required for a home loan. The population operates under one of three policy environments, a baseline scenario, a reduced deposit scenario (RD) and an early withdrawal scenario (EW). The only difference between the baseline and the RD scenarios is the minimum deposit required. Hence both these scenarios are characterized under Option 1. The EW scenario allows for early pension withdrawal for homeownership, that is, households are permitted to combine savings with part of their pension account. This leads to changes in the dynamics, and hence is characterized as Option 2.
The analysis compares how the population fares under each policy framework, as well as under a reduction in the required minimum deposit.

Model variables depend on several household-specific characteristics, including survival status, household structure (single vs.\ couple), income percentile (e.g.\ affecting pension contributions), and target property price (e.g.\ determining the required deposit threshold). Additionally, the policy governing pension withdrawals for housing impacts broader economic variables such as house price and rent inflation, as well as pension account balance. For brevity, these dependencies are not explicitly reflected in the notation within this section, except in the definition of the number of homeowners.

\subsection{Model dynamics}\label{Sec:Model2}
At time $t=0$, all households plan for a specific target property characterized by certain attributes (e.g.\ size, location) and an initial price $P(0)$. The price of this target property evolves over time, with value at time $t$ denoted by $P(t)$. Households maintain a fixed preference for properties with these characteristics.

The balance at time $t$ on the savings and pension accounts are denoted by $A(t)$ and $F(t)$, respectively. The savings account balance earns a return at rate $r_A(t-1)$ over $[t-1,t)$ and is affected by disposable income $I(t)$ (net of pension contributions and income taxes), non-housing consumption $C(t)$, and housing consumption $H(t)$, all associated with period $[t,t+1)$. The pension account balance earns a return at rate $r_F(t-1)$ over $[t-1,t)$, and increases through pension contributions $M(t)$ over $[t,t+1)$ before retirement, or decreases by the pension benefit withdrawal $B(t)$ over $[t,t+1)$ after retirement.  Taxes on returns over $[t-1,t)$ on the savings and pension account are denoted by $\tau_A(t)$ and $\tau_F(t)$, and are assumed to be paid at time $t$.

The savings and pension account balances are also reduced by withdrawal amounts $D_A(t)\geq0$ and $D_F(t)\geq0$, respectively. These amounts are used to finance the home purchase and are equal to zero for all $t$ except at the purchase time $T_p$. At $t=T_p$, the combined withdrawal must be sufficient to cover the required home loan deposit, property transfer tax, and an additional buffer for other purchase-related expenses. Specifically, for $t\neq T_p$, $D_A(t)=D_F(t)=0$, while at $t=T_p$:
\begin{equation}\label{Eq_2}
	D_A(T_p) + D_F(T_p)=\left(\delta+\epsilon\right)P(T_p)+\tau_P(T_p),
\end{equation}
where $\delta$ and $\epsilon$ represent the required deposit and buffer as fractions of the property price $P(T_p)$, and $\tau_P(T_p)$ denotes the property transfer tax. Under Option 1, where pension withdrawals are not allowed, $D_F(T_p)=0$, meaning that the entire amount must be withdrawn from savings. Under Option 2, which allows early pension withdrawal for homeownership, $D_A(T_p)$ and $D_F(T_p)$ must satisfy specific policy constraints. The exact expressions for these withdrawals under Option 2 are provided in Section \ref{Sec:Model3}, as they are closely tied to the definition of the purchase time.

The savings and pension account balances at time $t$ are given by:
\begin{eqnarray}\label{Eq_8}A(t) &=&	A_{acc}(t) + I(t) - C(t) - H(t)-D_A(t),\\
	\label{Eq_9} F(t) &=& F_{acc}(t) + M(t) - B(t)  - D_F(t),
\end{eqnarray}
where $A_{acc}(t)$ and $F_{acc}(t)$ represent the accumulated savings and pension accounts net of previous period's taxes:
\begin{eqnarray}
	A_{acc}(t) &=&	A(t-1)(1+r_A(t-1))-\tau_A(t),\label{Eq_10} \\
	F_{acc}(t) &=& F(t-1)(1+r_F(t-1))-\tau_F(t),\label{Eq_11}
\end{eqnarray}
with initial values $A_{acc}(0)$ and $F_{acc}(0)$.


The disposable income $I(t)$ is defined as follows:
\begin{equation}\label{Eq_6}
	I(t) =
	\left\{\begin{array}{ll}
		S(t)-\tau_S(t),& \text{for } t< T_r,\\
		B(t) + G(t) ,& \text{for } t\geq T_r,\end{array}\right.
\end{equation}
where $S(t)$ is the gross income earned over $[t,t+1)$ after pension contributions, $\tau_S(t)$ is the income tax assumed to be deducted directly from salary, and $G(t)$ is the guaranteed state pension received after retirement. Thus, pre-retirement income is the net earnings after tax and pension contribution, while post-retirement income consists of pension withdrawals $B(t)$ and state benefits $G(t)$ (e.g.\ social security).

Pension contribution $M(t)$ is given by:
\begin{equation}\label{x6}
		M(t) =
	\left\{\begin{array}{ll}
		(1-\tau_{\gamma}(t))\gamma S(t),& \text{for } t< T_r,\\
		0 ,& \text{for } t\geq T_r,\end{array}\right.
\end{equation}
where $\gamma$ is the contribution rate to the pension fund and $\tau_{\gamma}(t)$ is the tax rate on contributions.

The housing consumption $H(t)$ evolves as follows:
\begin{equation}\label{Eq_5}
	H(t)=\left\{\begin{array}{ll}
		R(t), & \text{ for } t <T_p,\\
		\mu(t)+ \pi(t), & \text{ for } T_p\leq t< T_p+T_l,\\
		\mu(t), & \text{ for } t\geq T_l + T_p,
	\end{array}\right.
\end{equation}
where $R(t)$ is the rent paid before property purchase, $\mu(t)$ captures ongoing property-related costs (e.g.\ maintenance, local property tax) after purchase, and $\pi(t)$ are the mortgage repayments made at the beginning of each period until the loan is repaid by term $T_l$.

The mortgage repayments follow standard actuarial formulas for loans payable in advance:
\begin{equation}\label{Eq_3}
	\pi(t)=\frac{L(t)r_B(t)}{1-\left(1+r_B(t)\right)^{-(T_l+T_p-t)}}, \quad \text{ for } t=T_p,...,T_p+T_l-1,
\end{equation}
where $r_B(t)$ is the time-varying borrowing rate, and $L(t)$ is the outstanding loan balance:
\begin{equation}\label{Eq_4}
	L(t)=\left\{\begin{array}{ll}
		0, & \text{ for } t <T_p,\\
		(1-\delta)P(T_p), & \text{ for } t=T_p,\\
		\left(L(t-1)-\pi(t-1)\right)(1+r_B(t-1)), & \text{ for } T_p+1\leq t\leq T_p+T_l-1.\\
		0,& \text{ for } t\geq T_p+T_l.
	\end{array}\right.
\end{equation}
Specifically, the household has no mortgage debt before property purchase. At the purchase time $T_p$, the outstanding loan equals the property price minus the deposit. During the loan term, the balance decreases with repayments and increases with interest. After $T_p+T_l$, the loan is fully repaid.

For homeowners, it is assumed that at the first time $t^{\star}$ such that  $A_{acc}(t^{\star}) \leq 0$ and $t^{\star} \geq T_p$, the household’s property is liquidated, meaning that accumulated savings are increased by $P(t^{\star}) - L(t^{\star})$, where the property is sold at the prevailing market price and net of the outstanding loan balance. In this case, the property purchase status and time are reset, and the household starts paying rent again, and may re-purchase a property later on. 

The default time $T_d$ is defined as the first time $t$ such that $A(t) \leq 0$ and the household does not own property. Upon default, the household exits the model, and its default status is set to  $d(t) = 1$ for $t \geq T_d$.

Before defining the model’s output, it is relevant to note that temporary income loss due to unemployment or disability is not explicitly modeled. However, many pension funds offer income protection and total permanent disability insurance, which offset this omission. Additionally, for simplicity, the analysis excludes death benefits that can be provided by pension funds.

\subsection{Model output}\label{Sec:Model3}

The primary output of the life-cycle model is the purchase time, which is the time (or period $[T_p,T_p+1)$) when the household acquires a property. Under Option 1, households rely solely on accumulated savings, and the purchase occurs at the first time $t$ when the accumulated savings balance $A_{acc}(t)$ meets or exceeds the required amount $(\delta+\epsilon)P(t)+\tau_P(t)$. Under Option 2, households can supplement their savings with withdrawals from their pension account, and the purchase occurs at the first time $t$ when an admissible combination of accumulated savings $A_{acc}(t)$ and accumulated pension account $F_{acc}(t)$ meets or exceeds the required amount. Under either option, the household's income net of taxes and consumption must exceed the repayment at the time of purchase. 

In Option 2 considered in this paper, the policy parameters are defined as follows: a minimum proportion $\alpha$ of the property price must be covered by savings; a maximum allowable withdrawal proportion $\beta$ from the pension account; and an absolute cap $F^{\max}$ on pension withdrawals. Additionally, pension withdrawals can only be used toward the home loan deposit, meaning that savings must still meet the remaining deposit requirement. These policy parameters align with the regulatory framework proposed by the Australian Coalition, but they can be tuned to match other more general policies. To simplify the decision-making process, the model assumes that households withdraw the full permissible amount from their pension accounts and use savings to cover any remaining shortfall. 

Therefore, the purchase time is the first time when two conditions are met. First, the household’s accumulated savings must meet the deposit affordability threshold, that is, $A_{acc}(t) \geq (\delta+\epsilon)P(t)+\tau_P(t)-D_F(t)$). Second, the household’s disposable income must meet the repayment affordability threshold, that is, $I(t)\geq \pi^{ref}(t)$ where $\pi^{ref}(t)=\frac{(1-\delta)P(t)r_B(t)}{1-\left(1+r_B(t)\right)^{-T_l}}$ is the first repayment that would apply in case of purchase at time $t$. Thus, $T_p$ under either policy scenario satisfies:
\begin{equation}\label{Eq_13}
	T_p = \inf\left\{t\leq \min\{T_\sigma,T_d\} \text{ } \mid \text{ } I(t)\geq \pi^{ref}(t) \text{ and } A_{acc}(t) \geq (\delta+\epsilon)P(t)+\tau_P(t)-D_F(t)\right\}.
\end{equation}
The withdrawals from savings and pension accounts at time $T_p$ are given by:
\begin{eqnarray}
	D_F(T_p)&=&\min\left\{\beta F_{acc}(T_p),F^{\max},(\delta-\alpha)P(T_p) \right\},\label{Eq_14}\\
   D_A(T_p)&=&(\delta+\epsilon)P(T_p)+\tau_P(T_p)- D_F(T_p).\label{Eq_15}\end{eqnarray}
Under Option 2, the accumulated savings must exceed a \textit{reduced} required amount, with a reduction of $D_F(T_p)$ used from the pension account. The reduction is equal to the proportion $\beta F_{acc}(t)$, but cannot exceed the maximum withdrawable amount $F^{\max}$. Furthermore, the reduction can only be used towards the deposit while allowing for a minimum $\alpha P(t)$ from the savings, meaning that the reduction cannot exceed $(\delta-\alpha)P(t)$. Notice that in all cases, $D_F(T_p),D_A(T_p)\geq 0$ and $D_A(T_p)+D_F(T_p)=(\delta+\epsilon)P(T_p)+\tau_P(T_p)$. Additionally, under Option 1 where $\alpha=\beta=F^{\max}=0$, it follows that $D_F(T_p)=0$ and $D_A(T_p)=(\delta+\epsilon)P(T_p)+\tau_P(T_p)$.

A second key output of the model is the number of homeowners at time $t$, and is denoted by $N_H(t)$. Introducing household superscripts, the purchase time of household $i\in\mathcal{K}$ is $T_p^{(i)}$, where $\mathcal{K}$ may refer to the entire population or to a specific income percentile, and $N_H(t)$ is given by:
\begin{equation}\label{Eq_16}N_H(t)=\underset{i\in \mathcal{K}}{\sum}\sigma^{(i)}(t)(1-d^{(i)}(t))\mathbb{I}\left(T_p^{(i)}\leq t\right),\end{equation}
where $\mathbb{I}\left(T_p^{(i)}\geq t\right)$ equals 1 if household $i$ has purchased by time $t$. 

Table \ref{Table0} classifies all variables in the model into input or output variables. Input variables (e.g. tax rules, macroeconomic variables, financial market parameters) are exogenous determinants that may be fixed or modeled as stochastic processes estimated from historical data. These variables are specified in Section \ref{Section-Sim} for the Australian case. Output variables are determined endogenously based on the input variables (e.g. savings and pension account balances, disposable income, time of purchase).
\FloatBarrier
\begin{table}[!h]
	\centering
	\begin{tabular}{lll}
		\toprule
		\multicolumn{3}{l}{\textbf{\textit{Input variables}}}\\
		\multicolumn{3}{l}{\textit{Population variables}}\\
		&$T_{\sigma}$ & Time of exit due to death\\
		&$T_{d}$& Time of exit due to default \\
		\multicolumn{3}{l}{\textit{Tax variables}}\\		
		& $\tau_P(t)$ & Property transfer tax upon purchase  \\ 
		&$\tau_I(t)$ & Income tax  \\ 
		&$\tau_{\gamma}$ & Tax rate on employer's pension contributions  \\ 
		&$\tau_A(t)$, $\tau_F(t)$ & Taxes on savings and pension investment returns \\
		\multicolumn{3}{l}{\textit{Pension variables}}\\
		&$T_r$ & Time of retirement \\ 
		&$\gamma$ & Employer's mandatory pension contribution rate (proportion of gross salary) \\ 
		&$B(t)$&Pension withdrawal benefit \\
		&$G(t)$&Social Security benefit\\		
		\multicolumn{3}{l}{\textit{Households variables}}\\
		&$\delta$ & Home loan deposit rate \\ 
		&$\epsilon$ & Buffer rate for additional costs  \\ 
		&$T_l$ & Loan term \\
		&$\mu(t)$ & Property maintenance costs  \\ 
		&$C(t)$&Non-housing consumption\\		
		\multicolumn{3}{l}{\textit{Stochastic input variables}}\\		
		&$N(t)$ & Number of surviving households \\
		&$\sigma(t)$ & Survival status \\
		&$P(t)$ & Target property price \\ 
		&$R(t)$ & Rent \\ 
		&$S(t)$ & Gross salary  \\ 
		&$r_B(t)$ & Borrowing rate \\ 
		&$r_A(t)$ & Savings account return \\ 
		&$r_F(t)$ & Pension account return  \\ 
		
		\multicolumn{3}{l}{\textit{Pension withdrawal policy rules}}\\
		&$\alpha$& Minimum proportion of the property price financed using savings account\\
		&$\beta$& Maximum proportion withdrawn from the pension account\\
		&$F^{\max}$& Maximum amount withdrawn from the pension account\\\hline
		\multicolumn{3}{l}{\textbf{\textit{Output variables}}}\\
		&$T_p$ & Time of property purchase  \\ 
		&$N_H(t)$&Number of homeowners in the population \\
		&$d^{(i)}(t)$& Default status \\
		&$A(t)$, $F(t)$ & Savings and pension account balance \\ 
		&$D_A(t)$, $D_F(t)$ & Amounts withdrawn from savings and pension account at the purchase time  \\ 
		&$H(t)$ & Housing consumption  \\ 
		&$I(t)$ & Disposable income  \\
		&$M(t)$ & Mandatory employers pension contribution  \\
		&$\pi(t)$ & Loan repayment  \\ 
		&$L(t)$ & Outstanding loan debt  \\
		\bottomrule
	\end{tabular}
		\caption{\textbf{Summary of notations of the general life-cycle model.} \textit{ The top panel lists the input variables which may be fixed by law, or modeled as stochastic processes estimated from historical data. The bottom panel lists the model output variables, which are determined endogenously based on the input variables}.}
	\label{Table0}
\end{table}
\FloatBarrier
\section{Input variables for Australia and simulation design}\label{Section-Sim}
Australia was one of the first countries to introduce a mandatory DC system, the Superannuation Guarantee (SG), which requires employer contributions into privately managed accounts. These accounts supplement a means-tested public Age Pension \citep{2001Treasury,2024Mercer}. Superannuation assets now exceed AUD 3.5 trillion, making super the second-largest source of household wealth after owner-occupied housing \citep{2024Eslake}. Australia’s SG is most similar to Switzerland’s Occupational Pension scheme, Kazakhstan’s Unified Accumulative Pension Fund, and Singapore’s Central Provident Fund \citep{2002McCarthy,2023Akbobek,2024Mercer}. Other countries, such as New Zealand and the UK, have introduced opt-out systems with lower contribution rates, while Canada and the US have increasingly shifted toward voluntary DC schemes without making them fully mandatory.

The SG was designed with the assumption that most retirees would be homeowners. However, as in most countries, homeownership rates have declined, particularly among younger and lower-income households \citep{2010YatesBradbury,2021MckellInstitute,2024Eslake}. In response, Australian governments have introduced demand-side housing policies, such as stamp duty exemptions and first-home buyer incentives, which have drawn criticism for inflating prices rather than improving accessibility \citep{2021MckellInstitute,2023Agarwal}. Similar issues have been observed internationally in the context of Mortgage Interest Deductions (MID) and housing subsidies across Germany, the UK, and the US \citep{2006Vigdor,2013Bourassa,2020BergerTurnerZwick,2023Krolage,2024Carozzi}.

Two recent housing policy proposals in Australia are of interest: the Reduced Deposit (RD) policy and Early Withdrawal (EW) from superannuation accounts. The Labor Party, which won the election, proposed a RD policy that lowers the required deposit from the typical 20\% to 5\%, backed by a government guarantee. This guarantee removes the need for Lenders Mortgage Insurance, which is often required for loans with deposits under 20\%. EW, originally proposed by the Liberal-National Coalition in 2022 and reiterated during the 2025 campaign, allows eligible first-home buyers to withdraw 40\% of the superannuation balance to help cover their deposit. Households cannot withdraw more than AUD 50,000, and must contribute 5\% of the property price from their savings account.

These proposals are evaluated relative to a benchmark with a standard deposit requirement. In the notations of Section \ref{Section-model}, the benchmark corresponds to Option 1 with $\delta=20\%$, and $\alpha=\beta=F^{\max}=0$. The RD policy corresponds to Option 1 but with a reduced deposit $\delta = 5\%$. The EW policy corresponds to Option 2, with $\delta=20\%$, a minimum savings contribution of $\alpha = 5\%$ of the property price, a maximum withdrawal of $\beta = 40\%$, and an absolute withdrawal cap of $F^{\max} = 50{,}000$.

The rest of this section is structured as follows. Section \ref{Subsec:Super} provides additional background on these housing policy proposals and similar schemes in other countries. Section \ref{Subsec:input} outlines the fixed input parameters at $t = 0$, including tax, pension, and household assumptions based on the Australian context. It also presents the econometric framework for simulating stochastic variables. Section \ref{Subsec:sim} details the simulation procedure. Section \ref{Subsec:metrics} introduces the outcome metrics used to assess the policy effects.

\subsection{Policy context}\label{Subsec:Super}
Both EW and RD policies aim to improve homeownership accessibility by lowering the upfront savings threshold to 5\% of the property price. However, they differ in their mechanics and distributional implications. EW allows individuals to draw on their own retirement savings, raising concerns about reduced pension balances. Yet its effect on retirement security is ambiguous, as converting super into housing may reduce post-retirement housing expenses \citep{2023Hand}. RD, by contrast, avoids pension withdrawals but increases loan sizes, shifting the constraint from savings to income.

Both policies can reinforce inequality by benefiting those already better positioned to buy. Under EW, high-income earners accumulate pension faster and can meet the 5\% cash requirement sooner. Low-income earners may lack sufficient balances and be priced out. Under RD, high-income households are less likely to be bound by income-based lending constraints. Lower-income earners remain constrained by repayment limits and must accumulate more than 5\% to meet income affordability thresholds, which becomes increasingly difficult as prices rise.

EW has faced strong criticism from economists and policy commentators, who argue that it inflates prices and undermines retirement adequacy \citep{2025SMC}. RD has attracted less scrutiny, possibly due to its framing as a cost-reducing measure. Internationally, similar policies exist. Countries like Singapore, Switzerland, and Kazakhstan allow EW for housing under mandatory DC schemes, while Canada, the US, and New Zealand permit restricted access. Yet few studies quantify their welfare impacts. Notably, under EW, \cite{2021MckellInstitute} project minimal homeownership gains and rising prices in Australia, while \cite{2023Akbobek} find negligible price effects in Kazakhstan.

\subsection{Input variables}\label{Subsec:input}
All monetary figures are in Australian dollars (AUD), with AUD 1 $\approx$ USD 0.60 at the time of writing. Time $t$ is measured in quarters to match the data frequency. Each household enters the model at age 25 ($t = 0$). Households start with no initial assets ($A_{acc}(0) = F_{acc}(0) = 0$) and no inheritance or external financial support. The model tracks their wealth accumulation and housing decisions over time. While household composition (e.g. singles vs. couples) affects the level of income and consumption, what substantially matters for affordability when $A_{acc}(0) = F_{acc}(0) = 0$ is the ratio of house prices to disposable income. To simplify the analysis, all households are modelled as singles, while differences in affordability are captured by varying the house-price-to-income ratio through two market settings: an equal-affordability market and a supply-constrained market, as detailed later in this section.

\subsubsection{Constant input variables: tax rules, pension rules, and household variables}\label{Subsec:tax}

The five tax functions $\tau_P(t)$, $\tau_I(t)$, $\tau_A(t)$, $\tau_F(t)$ and $\tau_{\gamma}(t)$ are based on rates and thresholds from the Victoria's State Revenue Office (\href{https://www.sro.vic.gov.au/rates-taxes-duties-and-levies/general-land-transfer-duty-property-current-rates}{link}) and the Australian Taxation Office (ATO) (\href{https://www.ato.gov.au/tax-rates-and-codes/tax-rates-australian-residents\#ato-Australianresidenttaxrates2020to2025}{link}). First-home buyer concessions and exemptions are excluded from $\tau_P(t)$ as they are temporary and subject to policy changes. Australian income tax applies to both salary and investment returns and operates on an annual basis while the analysis is conducted quarterly. Two simplifications are made to avoid unnecessary end-of-year adjustments. First, $\tau_I(t)$ is applied only to salary with official tax thresholds divided by four to approximate quarterly taxation. Second, $\tau_A(t)$ is adjusted separately to account for savings returns. Appendix \ref{Appendix:taxes} provides details on the tax functions and their numerical values.

The pension-related variables $T_r$, $\gamma$, $B(t)$, and $G(t)$ are sourced from the Australian Taxation Office (ATO) (\href{https://www.ato.gov.au/tax-rates-and-codes/key-superannuation-rates-and-thresholds}{link}) and Services Australia (\href{https://www.servicesaustralia.gov.au/age-pension}{link}). Retirement age is set at 67 ($T_r = 168$ quarters), which is the official retirement age in Australia. The superannuation contribution rate is fixed at $\gamma = 12\%$, in line with the rate in force since July 2025. While pension withdrawals $B(t)$ are discretionary, they are assumed to follow the ATO’s minimum required withdrawal rates for simplicity. These rates are adjusted to a quarterly basis by dividing the annual values by four. The state-guaranteed pension $G(t)$ comprises a base amount, supplements, and rental assistance. Eligibility and benefit levels depend on income and asset tests, with different thresholds for homeowners and non-homeowners. Detailed formulas and parameter values for the pension-related variables are provided in Appendix \ref{Appendix:pension}.

The loan term is fixed at 30 years ($T_l = 120$  quarters). The buffer $\epsilon$ is fixed at $1\%$ of the property purchase price, and is assumed to cover costs such as legal fees, mortgage fees and furniture purchase. Government support schemes, such as the First Home Guarantee and First Home Owner Grant, are temporary measures and are not incorporated into the analysis.

The quarterly property maintenance cost $\mu_p(t)$ depends on factors such as location, type, age, and size of the property, as well as household behavior. For simplicity, it is assumed to be 0.25\% of the property price $P(t)$, following guidelines from the Home Owners Association (\href{https://www.homeownersassociation.com.au/how-to-estimate-your-annual-home-maintenance-costs/#:~:text=The%201%25%20rule%20serves%20as,and%20$20%2C000%20annually%20for%20upkeep.}{link}). 

Households allocate a proportion $\nu(t)$ of their disposable income net of housing to non-housing consumption, such that $C(t) = \nu(t) I(t)$. The values of $\nu(t)$ are informed by estimates from an RBA study \citep{2014RBA}, which reports the share of disposable income allocated to total consumption for each age group, as well as the share of consumption devoted to non-housing items. Combining these figures yields an estimate of non-housing consumption as a proportion of income net of taxes and housing costs. Due to data limitations, $\nu(t)$ is assumed to be constant across income percentiles. The proportion $\nu(t)$ is set to 61.1\% for $t\in[0,39]$ (ages 25-34), 59.3\% for $t\in[40,79]$ (ages 35-44), 58.2\% for $t\in[80,119]$ (ages 45-54), 60.5\% for $t\in[120,159]$ (ages 55-64), 84.4\% for $t\geq 160$ (ages 65 and above). A minimum quarterly consumption level of \$1,200 (2025 dollars) is imposed and indexed to inflation, ensuring that basic consumption needs are met regardless of income.

\subsubsection{Stochastic variables}\label{Subsec:stochastic}
The study models a population of $N(0)=10,000$ individuals. At time 0, all households are initially alive and aged 25, i.e. $\sigma^{(i)}(0)=1$ for $i=1,...,N(0)$. The survival paths $\sigma^{(i)}(t)$ are simulated from Bernouilli distributions based on the latest 2020-24 male and female survival probabilities published by the Australian Government Actuary (\href{https://aga.gov.au/publications/life-tables/australian-life-tables-2020-22}{link}). The maximum age is 100.

The initial values of all macroeconomic and financial variables are fixed across households, except for initial income $S(0)$ and rent $R(0)$, which vary by income group. Income determines household percentile. Initial rent is different for each income group, and is set at 30\% of $S(0)$.

The evolution of economic variables is modelled using quarterly data from Q4 2002 to Q4 2022, chosen to align with the available property price series. The modelling approach broadly follows the cascade structure developed by \cite{Khemka2024}, with some important modifications.

First, this study uses more recent data and re-estimates the coefficients for key economic series, including the Consumer Price Index (CPI), real cash rate, borrowing spread, rent, and superannuation returns. Consistent with \cite{Khemka2024}, CPI growth and the cash rate (assumed to equal the savings rate $r_A(t)$) are modelled as AR(1) processes, while the spread between borrowing and savings rates (i.e. $r_B(t) - r_A(t)$) follows a random walk. Superannuation returns are modelled as the savings rate plus a 1\% constant equity risk premium, with random variation. The CPI index is normalised at 1 at time 0. Initial values are set to $r_A(0) = (1.0435)^{1/4} - 1$ (RBA cash rate, Dec 2024), $r_B(0) = (1.0599)^{1/4} - 1$ (lowest available borrowing rate, Dec 2024; \href{https://www.comparethemarket.com.au/}{link}), and $r_F(0) = r_A(0) + 0.01$. Rent growth is estimated via an AR(2) model with an error correction component applied to the first difference of the rental price index. Model coefficients are provided in Table \ref{tab3}.

Second, residual treatment deviates from \cite{Khemka2024}, who used parametric and independent residuals. Here, residuals for each variable are modelled empirically, transformed into pseudo-observations, and jointly modelled using a vine copula. This allows for contemporaneous dependence between shocks in macro-financial variables, and captures co-movement not explained by covariates.

The salary process also differs. Salaries evolve according to $S(t) = S(t-1) (1 + x_s(t))(1 + awe(t))$, where $awe(t)$ is wage growth, modelled as AR(1), and $x_s(t)$ captures the deterministic age profile, calibrated as $x_s(t) = \exp(0.033091 - 0.001462 \times t/4) - 1$ from HILDA survey data in \cite{2021khemka}. Initial salaries $S(0)$ reflect five interpolated income percentiles from the ABS census data for age 25: \$3,250, \$7,937.50, \$12,625, \$17,312.50, and \$22,000. For reference, the 2024 ABS quarterly median income across the entire population is \$18,148 (i.e.\ \$1,396 per week). At age 25, only the top income group exceeds this median. Ignoring wage inflation $awe(t)$, the second group surpasses the median after 2 quarters, the third after 3 years, the fourth after 6.75 years, and the fifth after 15.75 years.

The most significant modelling difference from \cite{Khemka2024} lies in the treatment of property prices. In this paper, the property price index explicitly incorporates endogenous demand from the life-cycle model (Section \ref{Section-model}) as a covariate. This interaction is crucial because the housing policies considered in this paper are expected to increase home affordability in the short-run, leading to more purchases, and in turn, higher demand that inflates property prices.

Specifically, at each time $t$, for given realizations of all variables, the housing demand growth $y_d(t)$ is determined endogenously as:
$$y_d(t)=\log(1+\eta(t)) - \log(1+\eta(t-1)),$$
with $y_d(0)=0$, where $\eta(t)=\max(0,N_H(t)-N_H(t-1))$ is the number of new homeowners over $[t-1,t)$, and $N_H(t)$ is defined in \eqref{Eq_16}. The variable $\eta(t)$ is restricted to be non-negative to avoid numerical issues arising in case of households' default.

To ensure empirical validity, the effect of $y_d$ on the property price index is estimated using historical data that aligns with model-derived measurements. The time series of homeownership rates $\eta(t)$ is constructed using ABS data on the number of new home loans, measured using the sum of Victoria’s quarterly data (June 2002 - September 2024) for owner-occupied construction, newly built purchases, and existing dwelling purchases. Property price growth $y_p$, which is obtained as  $y_p(t)=\log(P(t))-\log(P(t-1))$, is modeled as follows:
$$y_p(t) -0.01 = \underset{(\text{s.e. } 0.067) }{0.6988}\cdot\left(y_p(t-1)-0.01\right) + \underset{(\text{s.e. } 0.022)}{0.1293}\cdot y_d(t-1) + \epsilon_t^{(p)},$$
where $\epsilon_t^{(p)}$ is the error term. The equation's R$^2$ is 0.68. 

At time 0, the target property price $P(0)$ is fixed, and households are assumed to maintain a consistent preference for the type of property they aim to purchase. The value of $P(0)$ is income-dependent, and two housing market structures are considered.

The first is the \textit{equal-affordability market}, in which the target property price at time 0 is set to approximately 61.5 times the household’s quarterly income at age 25, i.e. 15.25 times the yearly income. This specification assumes that a suitably priced property exists for every income level. As a result, $P(0)$ ranges from \$200,000 to \$1,350,000 across the income distribution. However, this market is idealized and not fully representative of the Australian context, where properties priced around \$200,000 are rare, even in rural areas.

The second is the \textit{supply-constrained market}, where target property prices at time 0 are restricted to a fixed range between \$300,000 and \$1,000,000. This setup reflects more realistic supply-side constraints in the Australian housing market. Under this scenario, housing affordability varies across income groups: for high-income households, the target property price represents approximately 45.5 times their quarterly income, while for low-income households it is around 92 times. These values reflect the relative scarcity of low-priced housing and the greater feasibility for high-income earners to purchase homes in the upper range. For example, \$300,000 properties are generally limited to outer suburban or rural areas, while \$1,000,000 properties correspond to typical two- to three-bedroom apartments or townhouses in metropolitan locations.

\begin{table}[!h]
\resizebox{\textwidth}{!}{%
\begin{tabular}{ll}
\hline
\textbf{Variable}                                   & \textbf{Modelling Equation} \\
\hline
\textit{CPI growth}                   & $cpi(t) = 0.0065 + 0.2897 \cdot (cpi(t-1) - 0.0065) + \varepsilon^{(cpi)}_t$ \\
\addlinespace
\textit{AWE growth}                  & $awe(t) = 0.0021 + 0.5716 \cdot awe(t-1) + 0.2500 \cdot cpi(t-1) + \varepsilon^{(awe)}_t$ \\
\addlinespace
\textit{Real cash rate}            & $rr_A(t) = 0.6080 \cdot rr_A(t-1) + \varepsilon^{(r)}_t$ \\
\addlinespace
\textit{Nominal cash rate}          & $r_A(t) = \max\left(0, \exp(rr_A(t) + cpi(t)) - 1\right)$ \\
\addlinespace
\textit{Borrowing spread}            & $s_B(t) = \max(0.0034, \min(0.011, s_B(t-1) + \varepsilon^{(s)}_t))$ \\
\addlinespace
\textit{Borrowing rate}             & $r_B(t) = \max(0,\exp(s_B(t) + r_A(t)) - 1)$ \\
\addlinespace
\textit{Superannuation return}               & $r_F(t) = r_A(t) + 0.01 + \varepsilon^{(f)}_t$ \\
\addlinespace
\textit{Error correction term for rent} & $\text{ECM}_r(t-1) = R(t-1) - 0.1386 + 0.2336 \cdot P(t-1) - 1.0943 \cdot awe(t-1)$ \\
\addlinespace
\textit{Rent growth}               & $y_R(t) = 0.007 + 0.6533 \cdot (y_R(t-1) - 0.007) + 0.2832 \cdot (y_R(t-2) - 0.007)$\\
& \qquad \qquad $ - 0.0117 \cdot \text{ECM}_r(t-1) + \varepsilon^{(R)}_t$ \\
\addlinespace
\textit{Property price growth}      & $y_P(t) - 0.01 = 0.6988 \cdot (y_P(t-1) - 0.01) + 0.1293 \cdot y_d(t-1) + \varepsilon^{(P)}_t$ \\
\hline
\end{tabular}%
}
\caption{\small Macroeconomic and financial variable dynamics used in the simulation model. Equations follow the cascade framework of \cite{Khemka2024} with updated parameters based on Q4 2002–Q4 2022 Australian data.}
\label{tab3}
\end{table}

\subsection{Simulation design}\label{Subsec:sim}
The simulation framework consists of $m^{eco}=1,000$ economic scenarios, indexed by $m=1,...,m^{eco}$, where each economic scenario corresponds to a different set of dependent residuals of the macroeconomic and financial variable paths, generated from the estimated econometric model, and  applied uniformly across all households. Dependence between the residuals is incorporated using a vine copula.

To maintain heterogeneity across the population, each of the $m^{eco}$ scenarios also includes a household-specific error term applied to salary growth. The variance of the common salary residual is reduced by half, and the remaining half of the variance is attributed to this household-specific error term, which is sampled from the simulated residuals of the salary growth.

The initial values and simulated residuals plugged into the econometric model lead to the first-period realizations of property prices $P(1)$, rent  $R(1)$, income $S(1)$, borrowing rates $r_B(1)$, savings rates $r_A(1)$, and superannuation returns $r_F(1)$. These values, in combination with tax rules and pension regulations, are used within the life-cycle model to determine each household’s financial position and home purchase decision. This leads to the change in homeownership $\eta(1)$, which in turn leads to the demand growth $y_d(1)$. The simulation proceeds iteratively: in each period $t$, the macroeconomic and financial variables are updated using the estimated econometric model, incorporating the endogenous demand function $y_d(t-1)$. These updated values are fed into the life-cycle model to determine household decisions and extract $y_d(t)$ for the next period. This iterative process continues until all households have exited the model due to default or death.

The entire process is repeated for the three competing settings, Option 1 with required deposit of $\delta=20\%$ (i.e. benchmark), Option 2 where early pension withdrawal is allowed with required deposit $\delta=20\%$ (i.e. Coalition's proposal), and Option 1 with $\delta=5\%$ (i.e. Labor's proposal with reduced deposit). The same simulated residuals are used to ensure comparability.

Repeating the procedure for each economic scenario $m=1,...,m^{eco}$ yields realizations of the key output variables. Namely, for each household $i=1,...,N(0)$: the purchase time $T_p^{(i)}$, the savings account balance at retirement $A^{(i)}(T_r)$, the disposable income net of housing costs $I^{(i)}(t)-H^{(i)}(t)$ at each time $t\geq T_r$, the time death $T_{\sigma}^{(i)}$, the salary $S^{(i)}$, the guaranteed pension income $G^{(i)}(t)$, the property maintenance fees $\mu_p^{(j)}(t)$, and all tax functions. All these output variables are synthesized into the evaluation metrics described in the next subsection. 

Note that the time of default is not used in the metrics because none of the households in the population exits due to default. This is explained by two features of the model. The first feature is that property purchase is subject to the affordability constraint, where purchase is permitted only when disposable income exceeds repayments. The second is that non-housing consumption is expressed as a deterministic percentage of disposable income net of housing, with a lower bound. This means that households cannot spend more than their disposable income on non-housing consumption. This is in general not the case, as non-housing consumption of low-income groups tends to exceed disposable income. In particular, low-income groups in the model of the present analysis are more likely to access the property market than in real life.

\subsection{Evaluation metrics}\label{Subsec:metrics}
This subsection introduces seven metrics used to evaluate the two policies. The first three metrics assess the impact on household financial outcomes, measuring housing accessibility (probability of remaining a renter, and time of purchase), and retirement financial security. The next two metrics evaluate distributional effects across income percentiles, focusing on the Gini coefficients of the time of homeownership access and post-retirement financial security. The final two metrics take the government perspective by looking into the net present value of tax revenue minus government subsidies for both the federal and state (local) governments.

Throughout this section, all expectations and probabilities are taken over economic scenarios. Additionally, the tilde notation refers to quantities for the policy being evaluated, while the bar notation refers to those determined under the benchmark Option 1 with $\delta=20\%$. The notation $\mathcal{K}_k$ corresponds to the set of individuals $i$ in income group $k=1,...,5$ at time 0.

Housing accessibility is measured by $\Delta_a^{(k)}$ and $\Delta_p^{(k)}$, which represent the difference in the probability of purchase, and the expected purchase age (in years starting from age 25), respectively. When computed at the population level rather than for a specific income percentile, the notations $\Delta_a^{(\bullet)}$ and $\Delta_p^{(\bullet)}$ are used. The two metrics are defined as:
\begin{eqnarray*}
	\Delta_a^{(k)}&=&\frac{1}{|\mathcal{K}_k|}\sum_{i\in\mathcal{K}_k}\left(\mathbb{P}\left[\tilde{T}^{(i)}_p<T_{\sigma}^{(i)}\right] - \mathbb{P}\left[\bar{T}^{(i)}_p<T_{\sigma}^{(i)}\right]\right),\\
	\Delta_p^{(k)} &=& \frac{1}{|\mathcal{K}_k|}\sum_{i\in \mathcal{K}_k} \mathbb{E}\left[25 + \frac{\tilde{T}_p^{(i)}}{4}\right] - \mathbb{E}\left[25 + \frac{\bar{T}_p^{(i)}}{4}\right].
\end{eqnarray*}
The difference $\Delta_a^{(k)}$ evaluates the impact of a policy on the likelihood of purchasing, and positive values imply that the policy improves accessibility to homeownership. The difference $\Delta_p^{(k)}$ evaluates the impact of a policy on the age of purchase for those who purchase. Negative values indicates that the policy reduces the time to homeownership. Note that comparing the age of purchase across policies is challenging because some households may never purchase under one policy, leaving their purchase age undefined. Moreover, conditioning on purchase can be misleading, as a policy that enables more households to buy, albeit at older ages, may suggest an increase in conditional time of purchase despite being effective. To address this, households who do not purchase under a given policy are assigned a purchase time equal to the maximum survival horizon of 300 quarters. Under this convention, negative values of $\Delta^{(k)}_p$ indicate that the policy leads to earlier or more frequent purchases, capturing its positive impact.

The impact on retirement security is quantified using the relative difference $\Delta_s^{(k)}$ for income percentile $k$ and  $\Delta_s^{(\bullet)}$ at the population level. This metric measures the expected present value of post-retirement disposable income, net of housing consumption, along with accumulated savings at the time of retirement. It reflects a household’s ability to sustain consumption and maintain financial stability throughout retirement. Negative values indicate that retirement security deteriorates under a given policy. The relative difference $\Delta_s^{(k)}$ is defined as:
$$\Delta_s^{(k)} = \frac{1}{|\mathcal{K}_k|}\sum_{i\in \mathcal{K}_k}\mathbb{E}\left[ \frac{\tilde{\mathcal{I}}^{(i)}}{\bar{\mathcal{I}}^{(i)}}-1\right],$$
where $\bar{\mathcal{I}}^{(i)}$ and $\tilde{\mathcal{I}}^{(i)}$ represent the post-retirement disposable income net of housing consumption of household $i$ under one of the two proposed policy and under the benchmark, respectively, with:
$$\mathcal{I}^{(i)}=A^{(i)}(T_r)+\sum_{t\geq T_r} \sigma^{(i)}(t)(1 - d^{(i)}(t))v(T_r,t)\left(I^{(i)}(t)- H^{(i)}(t)\right),$$
with $v(T_r,t)$ being the discounting factor from time $t$ to retirement time $T_r$, determined using the cash-rate and inflation. The income measure $\mathcal{I}^{(i)}$ allows for the savings account balance at the time of retirement ($A^{(i)}(T_r)$) and captures the effect of homeownership or renting on post-retirement income through the $H^{(i)}(t)$ term, while both the social security payments and the pension fund balance are implicitly reflected through the $I^{(i)}(t)$ term. The measure ignores bequest in the form of any unused funds in the pension fund and the housing wealth at death of the household.

The impact of the housing policies on inequality is assessed through two Gini-based metrics, one for purchase time inequality denoted by $\Delta_{\mathcal{G}|p}$, and another for retirement security inequality denoted by $\Delta_{\mathcal{G}|s}$. These metrics measure how a policy affects disparities in homeownership access and post-retirement financial stability, and they are defined as follows:
\begin{eqnarray*}
	\Delta_{\mathcal{G}|p}&=&\tilde{\mathcal{G}}_p - \bar{\mathcal{G}}_p,\\
	\Delta_{\mathcal{G}|s}&=&\tilde{\mathcal{G}}_s-\bar{\mathcal{G}}_s,
\end{eqnarray*}
where $\bar{\mathcal{G}}_p$ and  $\tilde{\mathcal{G}}_p$ denote the expected Gini coefficients for purchase time inequality under a housing policy and the benchmark, respectively, and $\bar{\mathcal{G}}_s$ and $\tilde{\mathcal{G}}_s$ represent the corresponding expected Gini coefficients for retirement security. Specifically:
\begin{eqnarray*}
	\mathcal{G}_p&=& \mathbb{E}\left[\left(\underset{i=1}{\overset{N(0)}{\sum}}\underset{i^{\prime}=1}{\overset{N(0)}{\sum}}|T_p^{(i)}-T_p^{(i^{\prime})}|\right)/\left(2N(0)\underset{i=1}{\overset{N(0)}{\sum}}T_p^{(i)}\right)\right],\\
	\mathcal{G}_s &=& \mathbb{E}\left[\left(\underset{i=1}{\overset{N(0)}{\sum}}\underset{i^{\prime}=1}{\overset{N(0)}{\sum}}|\mathcal{I}^{(i)}-\mathcal{I}^{(i^{\prime})}|\right)/\left(2N(0)\underset{i=1}{\overset{N(0)}{\sum}}\mathcal{I}^{(i)}\right)\right].
\end{eqnarray*}
The Gini coefficient for purchase time $\mathcal{G}_p$ measures the dispersion of home acquisition timing across individuals, while the Gini coefficient of post-retirement financial stability captures disparities in the present value of post-retirement disposable income net of housing costs. Negative values of $\Delta_{\mathcal{G}|p}$ and $\Delta_{\mathcal{G}|s}$ suggest a reduction in disparties due to the introduction of the housing policy under interest.

The last two metrics evaluate the effect of a housing policy on government finances by comparing the net expected present value (NPV) of government revenues and expenditures under both policy settings. Specifically, $\Delta_{Federal}$ and $\Delta_{Local}$ measure the impact on the federal and local (state/council) governments, respectively, such that:
\begin{eqnarray*}
	\Delta_{Federal}&=& \tilde{V}_{Federal}-\bar{V}_{Federal},\\
	\Delta_{Local}&=&\tilde{V}_{Local}-\bar{V}_{Local},
\end{eqnarray*}
where $\bar{V}_{Federal}$ and $\tilde{V}_{Federal}$ are the NPV's of the federal government, while $\bar{V}_{Local}$ and $\tilde{V}_{Local}$ are the corresponding NPV's of the local government. In this model, federal government income consists of income tax ($\tau_I$), taxes on savings and pension returns ($\tau_A$ and $\tau_F$), and taxes on employer superannuation contributions ($\tau_{\gamma} \gamma S(t)$). The federal government’s expenditure is the Age Pension payments ($G(t)$). For the local government, revenue comes from the property transfer tax ($\tau_P(T_p)$) at the time of purchase, as well as ongoing council rates, which are assumed to be 40\% of property maintenance costs $\mu_p(t)$ post-purchase. Thus, $V_{Federal}=\frac{1}{N(0)}\underset{i=1}{\overset{N(0)}{\sum}}\mathbb{E}\left[\mathcal{V}^{(i)}_{Federal}\right]$ and $V_{Local}=\frac{1}{N(0)}\underset{i=1}{\overset{N(0)}{\sum}}\mathbb{E}\left[\mathcal{V}^{(i)}_{Local}\right]$, with
\begin{eqnarray*}
	\mathcal{V}^{(i)}_{Federal}&=&\underset{t\geq 0}{\sum}\sigma^{(i)}(t)(1-d^{(i)}(t))v(0,t)\left(\tau_I^{(i)}(t)+\tau_A^{(i)}(t)+\tau_F^{(i)}(t)+\tau_{\gamma}\gamma S^{(i)}(t)\mathbb{I}\left[t< T_r\right] - G^{(i)}(t)\right),\\
	\mathcal{V}^{(i)}_{Local}&=&\underset{t\geq 0}{\sum}\sigma^{(i)}(t)(1-d^{(i)}(t))v(0,t)\left(\tau_P(t)\mathbb{I}\left[t=T_p^{(i)} \right]+ 0.4\mu_p(t)\mathbb{I}\left[t\geq T_p^{(i)}\right]\right).
\end{eqnarray*}
\normalsize
The effect on total government revenue is given by $\Delta_{government} = \Delta_{Federal}+\Delta_{Local}$.
\section{Results}\label{Sec:Results}

\subsection{Impact of policies on property price}

\begin{figure}[!h]
	\centering\includegraphics[scale=0.8]{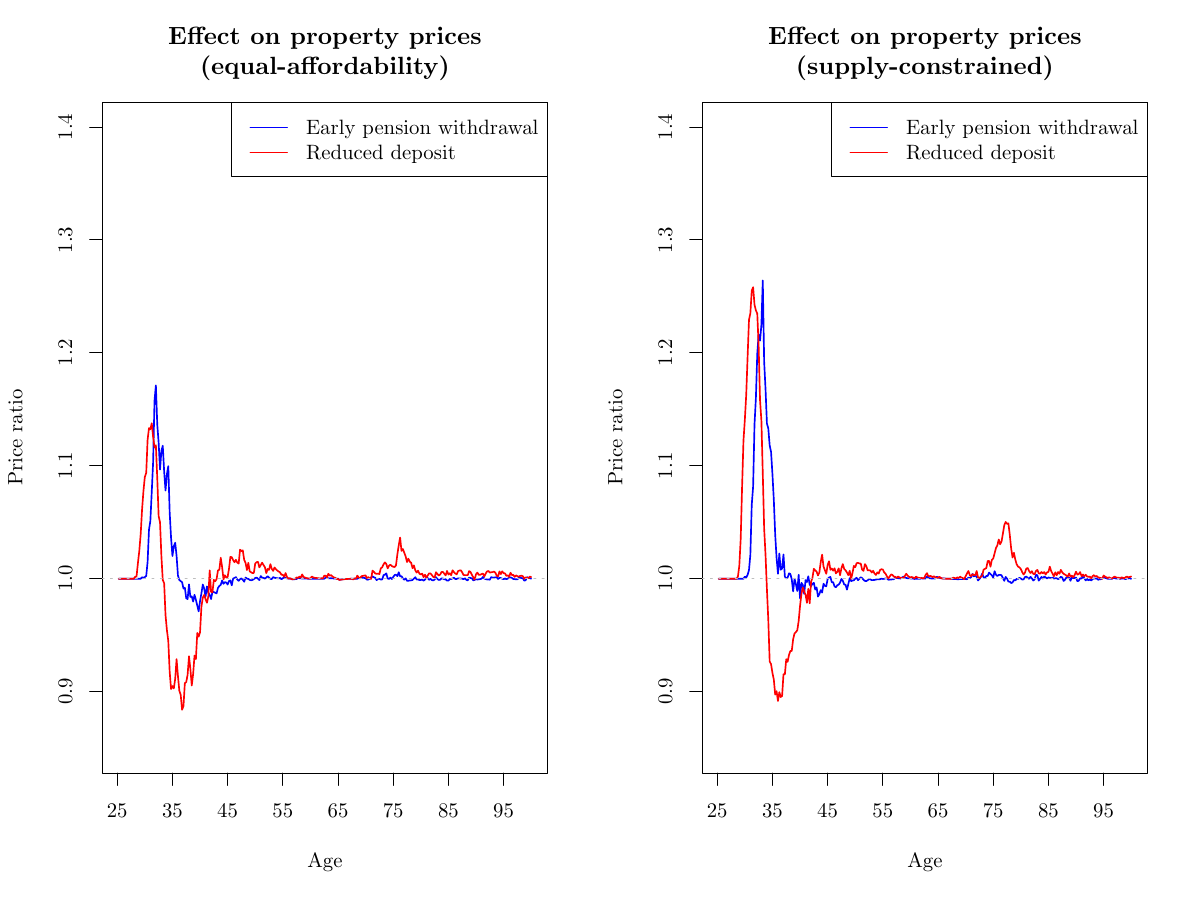}\caption{\small \textbf{Impact of housing policies on property price dynamics} -- \textit{Ratio of the property price index under each policy to the baseline scenario with no policy. Results are shown under two market assumptions: equal-affordability (left) and supply-constrained (right). The blue line corresponds to the Early Withdrawal policy, while the red line corresponds to the Reduced Deposit policy. Values above 1 mean that property price increase due to the introduction of the policy}.}\label{Figure1}
\end{figure}

Figure \ref{Figure1} displays the impact of the two housing policies on property prices, expressed as the ratio of the simulated property price index under each policy to the baseline scenario with no policy. Results are shown for both the equal-affordability (left panel) and supply-constrained market assumptions (right panel).

The estimates in Figure \ref{Figure1} are consistent in direction to the research of \cite{2025SMC}, but the magnitudes are higher. Specifically, \cite{2025SMC} suggests an average increase of 9\% across the capital cities, whereas Figure \ref{Figure1} suggests peaks around 20\%, which are especially more pronounced in the supply-constrained market.

Under both policy regimes, EW leads to a slightly higher peak in the property price index compared to RD. RD policy generates an earlier price response, due to its lower liquidity requirement. While both policies aim to reduce the deposit constraint, the EW policy requires a 5\% deposit from savings and 15\% from superannuation, whereas the RD policy requires only a 5\% deposit, with the remainder guaranteed by the government. Consequently, under the RD policy, buyers can enter the market more rapidly, which drives prices up sooner.

Under both policies and across both market assumptions, the property price index ratio converges to 1 in the long run. This indicates that the demand shock introduced by each policy is largely a temporal reallocation rather than a structural increase in total purchasing power. The implication is that, in the absence of ongoing new entrants, the long-term equilibrium price is not permanently shifted; only the timing of demand is altered. This provides a preliminary inference on the inefficiency of both housing policies.

Note that the absence of new entrants in the simulation design, Figure \ref{Figure1} exhibits two features that are unlikely in practice. The first feature is the temporary dip around age 35, which is due to the fact that once the initial wave of buyers enters the market earlier than they would under the baseline, the pool of new entrants dries up, creating downward pressure on prices. The second feature is the smaller peak is observed around retirement age under the EW policy, which appears because households who were unable to purchase earlier accumulate sufficient superannuation to enter the market later in life.

\subsection{Impact of policies on households and government income}
\begin{figure}[!h]
	\centering\includegraphics[scale=0.6]{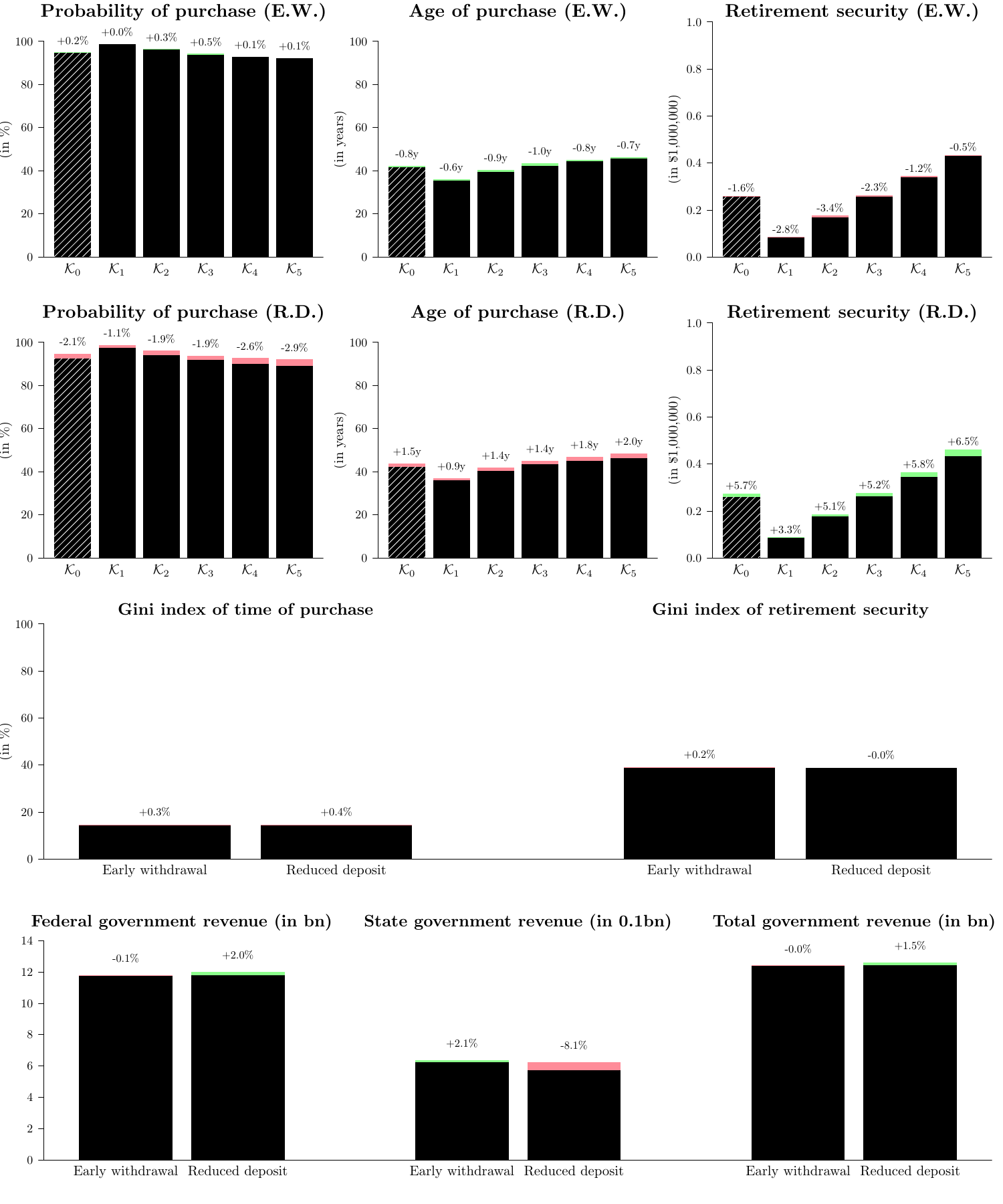}\caption{\small \textbf{Impact of housing policies on households and government income in the Equal-Affordability property market} -- \textit{
			Top two rows show effects on (i) the probability of purchase in percent, (ii) average age at purchase in years, and (iii) retirement financial security in million dollars, by income group $\mathcal{K}_1$ (lowest percentile) to $\mathcal{K}_5$ (highest percentile), with $\mathcal{K}_0$ (dashed bars) representing the full population. Results on the first row represent those of the early withdrawal policy (E.W.), and on the second row represent those of the reduced deposit policy (R.D.).
			Middle row reports the Gini indices in percent of purchase timing and retirement security.
			Bottom row presents the effect on the present value of federal government income in billion dollars (left), state government income in hundred million dollars (middle), and combined government income in billion dollars (right). For all panel, black bars represent baseline values under no policy, and coloured annotations reflect changes induced by the corresponding policy. Differences in green indicate a favourable outcome from the implementation of the policy, whereas differences in red indicate unfavourable one}.}\label{Figure2}
\end{figure}

\begin{figure}[!h]
	\centering\includegraphics[scale=0.6]{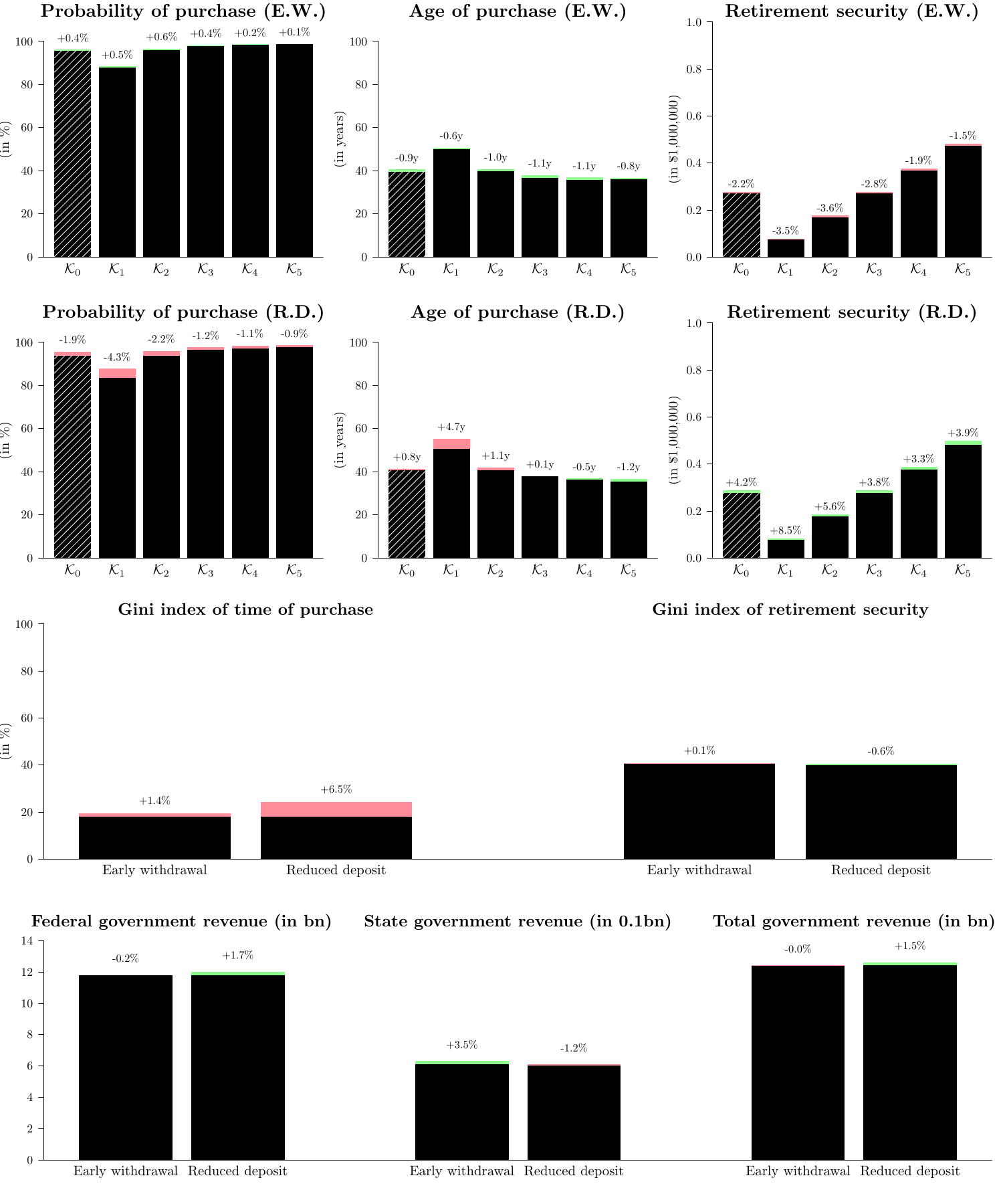}\caption{\small \textbf{Impact of housing policies on households and government income in the Supply-Constrained property market} -- \textit{
			Top two rows show effects on (i) the probability of purchase in percent, (ii) average age at purchase in years, and (iii) retirement financial security in million dollars, by income group $\mathcal{K}_1$ (lowest percentile) to $\mathcal{K}_5$ (highest percentile), with $\mathcal{K}_0$ (dashed bars) representing the full population. Results on the first row represent those of the early withdrawal policy (E.W.), and on the second row represent those of the reduced deposit policy (R.D.).
			Middle row reports the Gini indices in percent of purchase timing and retirement security.
			Bottom row presents the effect on the present value of federal government income in billion dollars (left), state government income in hundred million dollars (middle), and combined government income in billion dollars (right). For all panel, black bars represent baseline values under no policy, and coloured annotations reflect changes induced by the corresponding policy. Differences in green indicate a favourable outcome from the implementation of the policy, whereas differences in red indicate unfavourable one}.}\label{Figure3}
\end{figure}

The impact of the two housing policies on home purchase outcomes is presented in Figure \ref{Figure2} for the equal-affordability property market, and Figure \ref{Figure3} under the supply-constrained market. 
\FloatBarrier
\FloatBarrier

\FloatBarrier
\FloatBarrier

Beginning with the probability of purchase, the EW policy leads to a modest increase across the population, with gains below 1\% observed among all groups, mostly due to the fact that all households have an already high probability of purchase. In contrast, the RD policy consistently lowers that probability. In the equal-affordability market, the RD policy reduces accessibility for high-income earners more than low-income ones, whereas in the supply-constrained market, low-income earners experience the strongest negative effect with a 4.3\% decline in accessibility. This decline is due to the fact RD pushes prices up earlier than EW, and the income required to meet repayments is higher.

In terms of timing, EW shifts purchases earlier for all income groups, with lower gains for low-income groups (0.6 years), and higher gains for second and third income groups (1-1.1 years). RD leads to different effect. In the equal-affordability market, households from all income groups are expected to purchase about 0.9 to 2 years later, whereas in the supply-constrained market, high-income groups buy 1.2 years earlier and low-income groups buy 4.7 years later. These shifts in timing are reflected in the Gini index of the age of purchase. Specifically, both policies lead to marginal increases in inequality in the equal-affordability market, but in the supply-constrained market, RD leads to an increase of 6.5\% in the Gini index of accessibility, while EW leads to an increase of 1.4\%. The differing outcomes between the equal-affordability and supply-constrained markets indicate that these housing policies do not create inequality by themselves, but they tend to exacerbate inequalities that already exist in the housing market, with RD having the strongest effect.

Retirement security is adversely affected by the EW policy for all groups, especially in the supply-constrained markets, where losses range from -3.6\% for the second lowest-income households to -1.5\% for the highest group. This reflects the lower long-term returns from early property investment compared to retaining funds in superannuation. RD, by contrast, improves retirement outcomes across the board, with gains of 3.3–8.5\% depending on the income group and the property market setup. Gains in retirement security are largest for households who forego purchasing and benefit from compounding superannuation returns. None of the policies affects the Gini index of retirement security, with changes between -0.6\% and 0.2\% indicating that the gains do not significantly improve existing inequality. 

Under EW, federal government income remains largely unchanged, while state government revenue increases by approximately 2.1\% in the equal-affordability setting and 3.5\% in the supply-constrained setting. This is driven by earlier home purchases, which increases the present value of property transfer tax and longer collection of council rates. In contrast, RD leads to a small increase in federal government income around 2\%, reflecting stronger superannuation balances that reduce long-term Age Pension liabilities. However, state government revenue declines by 8.1\% under equal-affordability and 1.2\% under supply constrained, which is due to reduced transaction volumes and delayed purchases. Since the total government revenue in this setup is largely influenced by the revenue from the federal level, the combined figures for the total government revenue reflect the effect of the revenue from the federal level.

Overall, the RD policy undermines housing accessibility and exacerbates existing inequalities in supply-constrained markets, where it lowers the likelihood of purchase for low-income households and delays their market entry. In contrast, the EW policy facilitates earlier purchases across all income groups, though not uniformly. Its effectiveness in improving housing access comes at the cost of reduced retirement security for all households, given the lower long-term returns on housing relative to superannuation. Conversely, RD enhances retirement outcomes, but this improvement is incidental to its purpose and highlights a mismatch between intent and effect. From a public finance perspective, neither policy generates a clear fiscal benefit; EW marginally increases local government revenue via earlier stamp duty payments, while RD slightly improves federal government balances but reduces local government income. 

\subsection{Effect of price sensitivity to demand}
Sensitivity to demand was examined by varying the property-price growth coefficient in two directions, either a high-demand case (coefficient doubled) or a low-demand case (coefficient halved); see Table \ref{Table1} for the baseline parameter. Results are not reported because these changes do not materially alter the magnitude of policy effects under either market structure.

As expected, unreported figures show a larger surge of about 60\% when the coefficient is doubled and a smaller surge of about 10\% when it is halved. However, household and gorvernment revenue outcomes are largely unaffected. In the equal-affordability market, doubling or halving the coefficient changes purchase probability, retirement security, Gini indices, and government revenue by roughly 0.1\%, and shifts the purchase age by about 0.1 years. In the supply-constrained market, differences are similarly small when comparing low- and high-demand cases, except for retirement security. The response is mildly nonlinear, where both EW and RD improve when the coefficient is either doubled or halved, but the gains do not exceed 2\%. Overall, a stronger or weaker demand sensitivity shifts both the benchmark and the policy paths in the same direction, but leaves the relative policy effects essentially unchanged. 

\subsection{Effect of superannuation return}
Superannuation returns influence the model through two main channels. First, they affect the timing and likelihood of home purchase. Under the EW policy, lower returns reduce the future pension balance available for withdrawal, potentially delaying purchase. Under both EW and RD, super returns also affect purchase probability and timing for households that can afford buying until after retirement. Second, superannuation returns directly impact retirement financial security. In the baseline calibration, superannuation outperforms housing investment, meaning early purchase leads to forgone pension returns, which in turn reduces retirement security. This trade-off is most evident under the RD policy in the baseline, which delays or prevents purchase and thereby improves retirement outcomes.

The average superannuation return was reduced to match the average growth rate of property prices. Unreported results show that this adjustment had minimal impact on the property price path under RD but attenuated the price surge under EW. Specifically, whereas Figure \ref{Figure1} showed peak price increases of 18\% (equal-affordability) and 26\% (supply-constrained) under EW, these peaks declined to approximately 13\% and 23\%, respectively. In terms of household outcomes and government revenue, lowering the superannuation return led to negligible differences. Compared to the baseline, relative changes in purchase probability are below 1\% compared to the baseline, and average purchase ages are around $\pm$0.2 years. The impact on retirement financial security was more pronounced. Under EW, the effect turned slightly positive (compared to negative in the baseline), while under RD, the gains in retirement security roughly doubled.

\section{Policy alternatives}\label{Sec:Alternatives}
This section explores two alternative designs of the EW and RD policies. The first subsection examines the effects of restricting policy access to lower income households only. The second subsection analyzes extreme parameter settings for the EW policy: one in which households can withdraw up to 100\% of their superannuation balance (i.e. $\beta = 100\%$), and another where no savings contribution is required from the household (i.e. $\alpha = 0\%$), meaning the entire 40\% withdrawn amount can be used toward the deposit.

\subsection{Impact of restricted policies}
\begin{figure}[!h]
	\centering\includegraphics[scale=0.6]{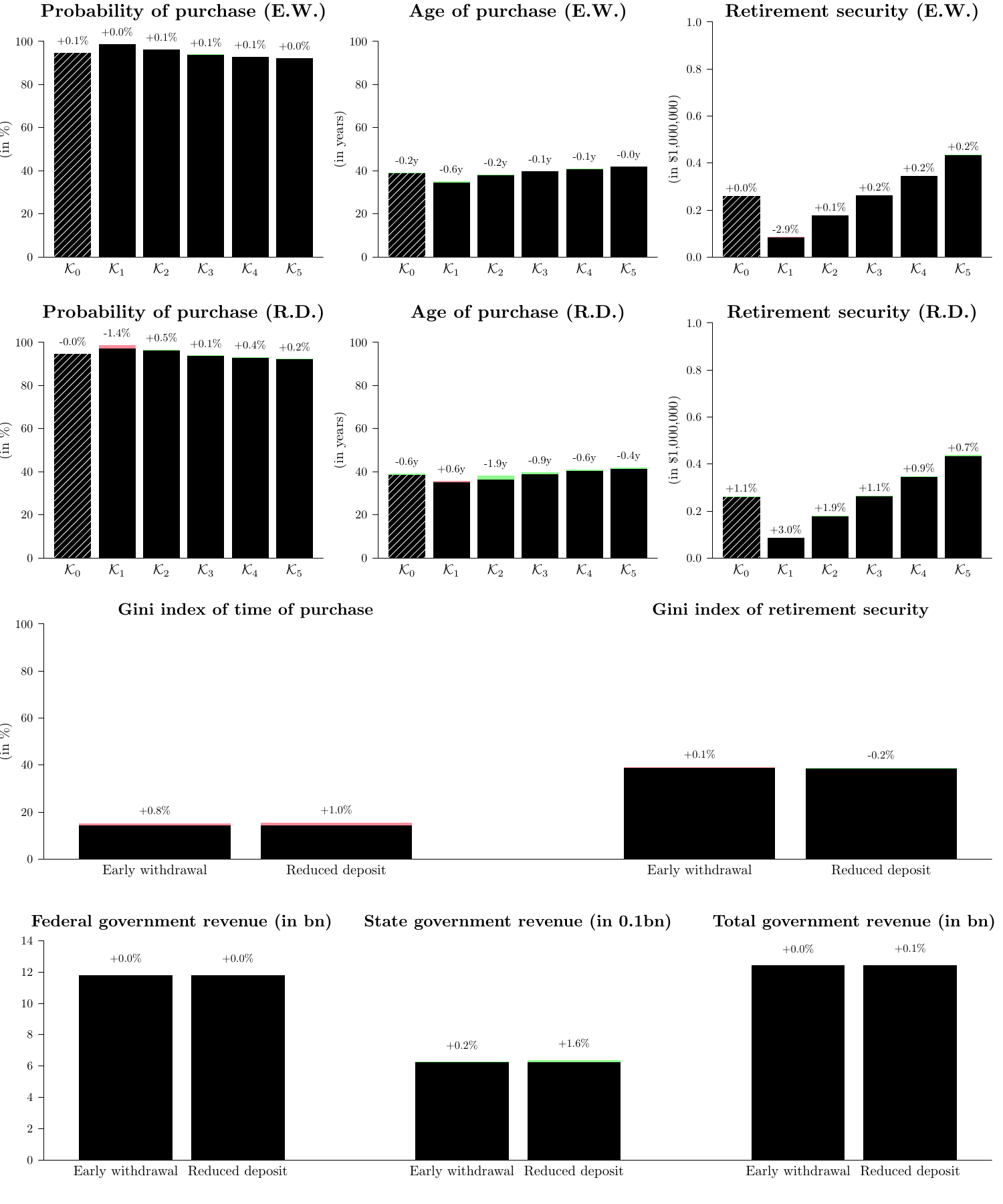}\caption{\small \textbf{Impact of restricted housing policies (access only to below all-ages population median income) on households and government income in the Equal-Affordability property market} -- \textit{
			Top two rows show effects on (i) the probability of purchase in percent, (ii) average age at purchase in years, and (iii) retirement financial security in million dollars, by income group $\mathcal{K}_1$ (lowest percentile) to $\mathcal{K}_5$ (highest percentile), with $\mathcal{K}_0$ (dashed bars) representing the full population. Results on the first row represent those of the early withdrawal policy (E.W.), and on the second row represent those of the reduced deposit policy (R.D.).
			Middle row reports the Gini indices in percent of purchase timing and retirement security.
			Bottom row presents the effect on the present value of federal government income in billion dollars (left), state government income in hundred million dollars (middle), and combined government income in billion dollars (right). For all panel, black bars represent baseline values under no policy, and coloured annotations reflect changes induced by the corresponding policy. Differences in green indicate a favourable outcome from the implementation of the policy, whereas differences in red indicate unfavourable one}.}\label{Figure4}
\end{figure}

\begin{figure}[!h]
	\centering\includegraphics[scale=0.6]{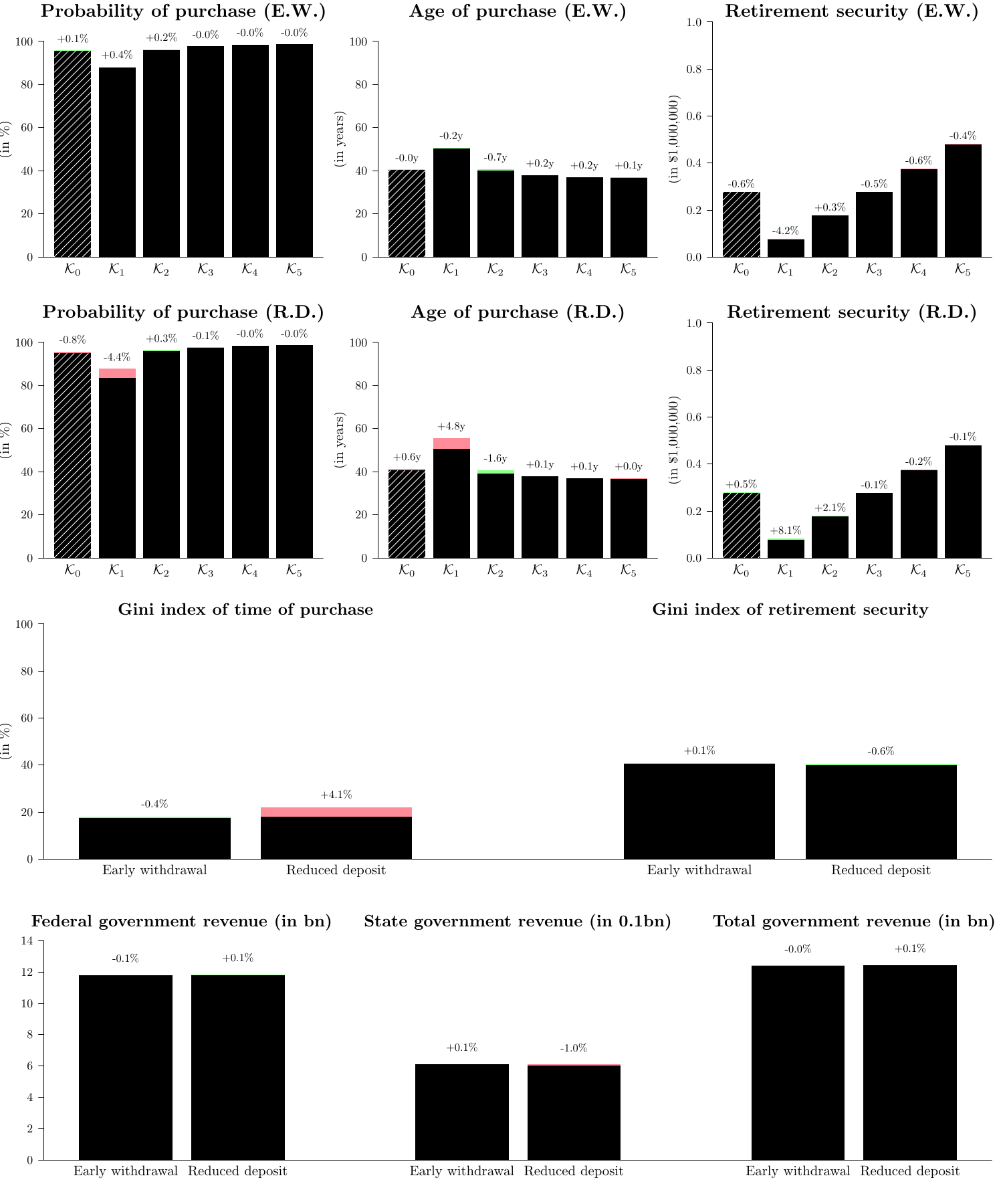}\caption{\small \textbf{Impact of restricted housing policies (access only to below all-ages population median income) on households and government income in the Supply-Constrained property market} -- \textit{
			Top two rows show effects on (i) the probability of purchase in percent, (ii) average age at purchase in years, and (iii) retirement financial security in million dollars, by income group $\mathcal{K}_1$ (lowest percentile) to $\mathcal{K}_5$ (highest percentile), with $\mathcal{K}_0$ (dashed bars) representing the full population. Results on the first row represent those of the early withdrawal policy (E.W.), and on the second row represent those of the reduced deposit policy (R.D.).
			Middle row reports the Gini indices in percent of purchase timing and retirement security.
			Bottom row presents the effect on the present value of federal government income in billion dollars (left), state government income in hundred million dollars (middle), and combined government income in billion dollars (right). For all panel, black bars represent baseline values under no policy, and coloured annotations reflect changes induced by the corresponding policy. Differences in green indicate a favourable outcome from the implementation of the policy, whereas differences in red indicate unfavourable one}.}\label{Figure5}
\end{figure}

Figures \ref{Figure4} and \ref{Figure5} report the main metrics for the equal-affordability and supply-constrained markets when policy access is limited to households below the all-age population median income. This restriction initially excludes only the top income percentile among 25-year-olds. Ignoring salary inflation $awe(t)$, the age profile implies that the lowest-income group at age 25 exceeds the population median by about age 40.

Restricting access below the median changes the distribution of outcomes only marginally and preserves the qualitative differences between policies. For EW, low income groups benefit as they would have under the unconstraint policy setup, although the gain in purchase age is lower in the supply-constrained market. Effects on the Gini indices and government revenue are not significant. For RD, the pattern also mirrors the universally accessible case, with attenuated effects for higher-income groups and amplified adverse effects for lower-income groups despite the policy being restricted. This occurs because the highest earners within the lower-income segment bid up property prices, leaving the least affluent households unable to meet the affordability constraint.

Overall, restricting EW to low income earners does not improve its efficiency, while a restricted RD does not improve accessibility for its intended beneficiaries and still increases inequality in purchase timing, especially under supply constraints. Unreported results with access restricted to the lowest 25th percentile show both policies leading to near-zero differences relative to the benchmark.

\FloatBarrier
\FloatBarrier
\FloatBarrier
\FloatBarrier

\subsection{Boundary cases of the EW policy}
The baseline EW design allows withdrawals up to $40\%$ of the pension balance, subject to an absolute cap of \$50{,}000, and requires a minimum $5\%$ contribution from the savings account ($\beta=40\%$, $F^{\max}=50{,}000$, $\alpha=5\%$). Two boundary cases are considered. In the first, households may withdraw their entire pension balance with the same $5\%$ savings contribution ($\beta=100\%$, $F^{\max}=\infty$, $\alpha=5\%$). In the second, withdrawals remain capped at $40\%$ but no savings contribution is required ($\beta=40\%$, $F^{\max}=50{,}000$, $\alpha=0\%$). Unreported results indicate that both boundary cases generate larger property price surges than the baseline, with surges reaching 40\% when $\beta=100\%$ in the supply-constrained market.

\begin{figure}[!h]
	\centering\includegraphics[scale=0.6]{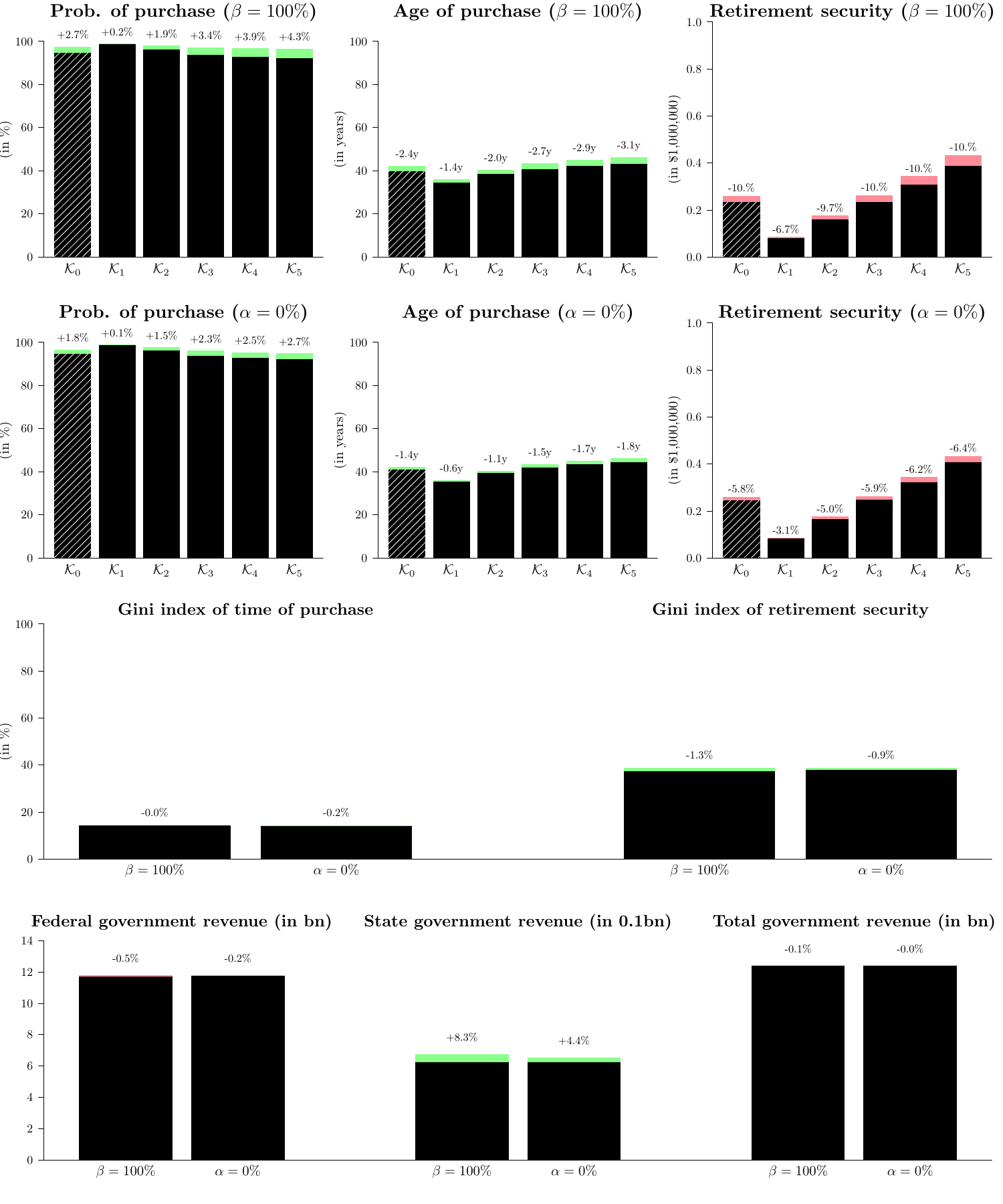}\caption{\small \textbf{Impact of EW on households and government income in the Equal-Affordability property market in the boundary cases where either $\beta=100\%$ and $\alpha=5\%$, or $\beta=40\%$ and $\alpha=0\%$} -- \textit{
			Top two rows show effects on (i) the probability of purchase in percent, (ii) average age at purchase in years, and (iii) retirement financial security in million dollars, by income group $\mathcal{K}_1$ (lowest percentile) to $\mathcal{K}_5$ (highest percentile), with $\mathcal{K}_0$ (dashed bars) representing the full population. Results on the first row represent those where $\beta=100\%$ and $\alpha=5\%$, and on the second row represent those where $\beta=40\%$ and $\alpha=0\%$.
			Middle row reports the Gini indices in percent of purchase timing and retirement security.
			Bottom row presents the effect on the present value of federal government income in billion dollars (left), state government income in hundred million dollars (middle), and combined government income in billion dollars (right). For all panel, black bars represent baseline values under no policy, and coloured annotations reflect changes induced by the corresponding policy. Differences in green indicate a favourable outcome from the implementation of the policy, whereas differences in red indicate unfavourable one}.}\label{Figure6}
\end{figure}

\begin{figure}[!h]
	\centering\includegraphics[scale=0.6]{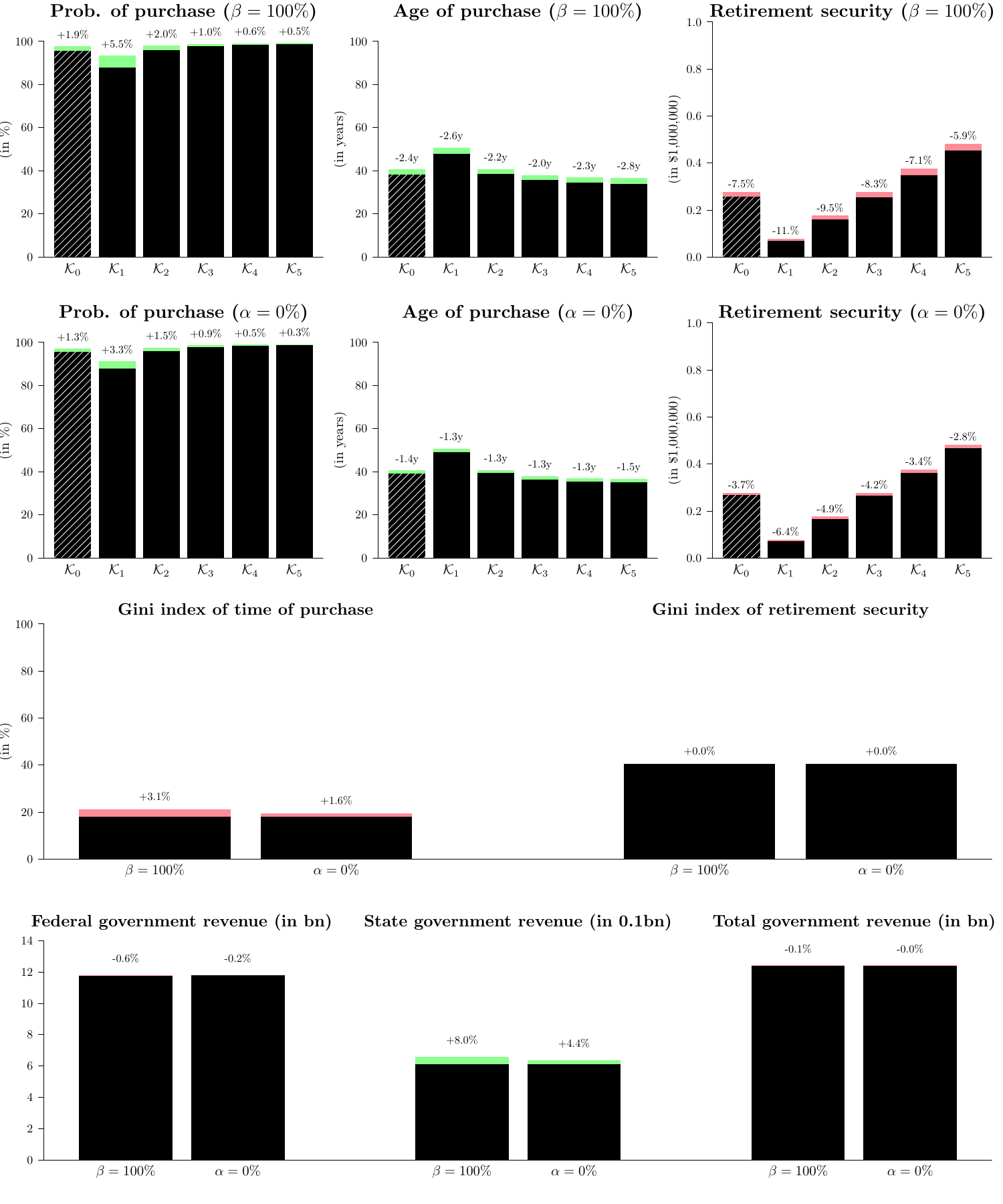}\caption{\small \textbf{Impact of EW on households and government income in the Supply-Constrained property market in the boundary cases where either $\beta=100\%$ and $\alpha=5\%$, or $\beta=40\%$ and $\alpha=0\%$} -- \textit{
			Top two rows show effects on (i) the probability of purchase in percent, (ii) average age at purchase in years, and (iii) retirement financial security in million dollars, by income group $\mathcal{K}_1$ (lowest percentile) to $\mathcal{K}_5$ (highest percentile), with $\mathcal{K}_0$ (dashed bars) representing the full population. Results on the first row represent those where $\beta=100\%$ and $\alpha=5\%$, and on the second row represent those where $\beta=40\%$ and $\alpha=0\%$.
			Middle row reports the Gini indices in percent of purchase timing and retirement security.
			Bottom row presents the effect on the present value of federal government income in billion dollars (left), state government income in hundred million dollars (middle), and combined government income in billion dollars (right). For all panel, black bars represent baseline values under no policy, and coloured annotations reflect changes induced by the corresponding policy. Differences in green indicate a favourable outcome from the implementation of the policy, whereas differences in red indicate unfavourable one}.}\label{Figure7}
\end{figure}

Figures \ref{Figure6} and \ref{Figure7} present the main metrics for the equal-affordability and supply-constrained markets under the two boundary cases. Both policies increase the probability of purchase relative to the baseline across market types. In the equal-affordability market, the rise is more pronounced among high-income households, whereas in the supply-constrained market, it is greater for low-income households, reaching up to $5.5\%$ when $\beta=100\%$. Boundary cases also lead to earlier purchases compared to the baseline, particularly for $\beta=100\%$. The largest shift in the equal-affordability market occurs among high-income households (3.1 years earlier), while in the supply-constrained market the improvement is relatively uniform across income groups (2.4 years earlier on average). The Gini index of purchase timing increases substantially in the supply-constrained market, but in this case, the inequality appears to benefit lower-income households.

Retirement security declines more sharply in the boundary cases, deteriorating by up to 11\% for low-income groups under $\beta=100\%$. By the return comparison logic, when pension returns exceed property price growth (as in the baseline), increasing purchases shifts wealth from the higher-return pension account to housing, thereby reducing retirement security. Because both boundary designs raise the probability of purchase, the deterioration is larger than under the baseline. The figures also show a pronounced rise in state government revenue from stamp duty, reflecting higher purchase activity. Total government revenue, however, remains essentially unchanged.

Note that consistent with Figures \ref{Figure4} and \ref{Figure5}, unreported simulations that restrict boundary-case access to low-income households yield only modest changes, indicating limited gains when targeting alone is used to mitigate adverse effects.

Overall, the boundary designs generate stronger effects than the baseline. Low-income households experience higher purchase probabilities, and all income groups are able to buy earlier. These results suggest that a social planner aiming to improve housing accessibility should permit full pension withdrawals, as this can increase market participation without worsening inequality among lower-income groups. However, such a strategy can undermine retirement security when pension returns substantially exceed property growth. This trade-off highlights the need for coordinated policy design that balances immediate housing accessibility with long-term financial sustainability, ensuring that short-term equity gains do not come at the cost of future welfare.

\FloatBarrier
\FloatBarrier
\FloatBarrier
\FloatBarrier
\section{Conclusion}\label{Sec:Conclusion}
This paper examines two housing policy proposals put forward in Australia in the lead-up to the 2025 federal election. Both policies aim to relax liquidity constraints on first-home buyers, thereby affecting the demand for housing. To evaluate their impact, a model is developed that links housing demand to housing prices for a cohort of 25-year-old with no initial savings, full employment throughout their working years, and no external financial support.

Both policies significantly increase property prices in the short term, with effects that would be amplified in a multi-cohort setting over the long term. RD is detrimental to housing accessibility of low-income groups, especially in supply-constrained markets. EW improves accessibility for all groups but raises a trade-off with retirement financial security when pension returns are substiantially high. Restricting access to below-median or bottom-quartile income groups does not materially alter the outcomes for unrestricted groups. Fiscal impacts are minor in both cases. The results are robust to doubling or halving the price sensitivity to demand. Boundary EW cases amplify purchase probabilities and price levels, and suggest that a social planner aiming at improving housing accessibility only should allow full pension withdrawal.

The paper also sharpens how price-inflating housing policies are interpreted, and derives broader results on liquidity shocks. Rising prices aren't always harmful. Both RD and EW raise prices in the short run, but RD delays or prevents access for lower-income households, whereas EW can significantly improve accessibility with little downside for inequality. Crucially, the results from RD's distributional effects are in large part driven by market structure, not by the price surge itself. In equal-affordability versus supply-constrained settings, low-income purchase timing shifts in opposite directions, indicating that pre-existing market disparities produce the unequal outcomes.

Future research could incorporate supply-side interventions from international experience, allowing for a deeper understanding of how to mitigate inequality in market structure. A further extension is to model multiple cohorts of new entrants so that feedback from repeated policy exposure, cross-cohort price dynamics, and intergenerational inequalities can be quantified.

\bibliographystyle{agsm}
\bibliography{Refs}

\newpage
\appendix
\renewcommand{\theequation}{\thesection.\arabic{equation}}

\section{Appendix}\label{App:}

\subsection{More details on tax rules}\label{Appendix:taxes}
The property transfer tax $\tau_P(t)$ follows the rates applicable in the state of Victoria and is defined as:
\begin{equation}\label{Eq3_1}
	\tau_P\left(t\right)=\left\{\begin{array}{ll}
			1.4\%  P(t), & \text{if } P(t) \leq  25,000, \\
			350 + 2.4\% \left(P(t) - 25,000\right), & \text{if } 25,000 < P(t) \leq 130,000, \\
			2,870+  6\% \left(P(t) - 130,000\right) , & \text{if } 130,000 < P(t) \leq 960,000,\\
			5.5\% P(t)  , & \text{if } 960,000 < P(t) \leq 2,000,000,\\
			110,000 + 6.5\% \left(P(t) - 2,000,000\right)  , & \text{if } P(t) > 2,000,000.
		\end{array}\right.
	\end{equation}
	
	In Australia, income tax applies to both gross salary and returns on savings, inherently linking the functions $\tau_I(t)$ and $\tau_A(t)$. The total income tax is given by:
	\begin{equation}
		\label{Eq3_2}
		\tau_S(t) + \tau_A(t) =  \left\{
		\begin{array}{ll}f_{\text{income tax}}\left(S(t)+r_A(t-1)A(t-1)\right), & \text{for } t < T_r,\\
			f_{\text{income tax}}\left(r_A(t-1)A(t-1)\right),& \text{for } t \geq T_r,
			\end{array}\right.
	\end{equation}
	which reflects the fact that taxable income after retirement comes from investment returns only.
	
	Since income tax is assessed annually, but this study operates on a quarterly basis, tax thresholds in the function $f_{\text{income tax}}$ are divided by four:
	$$
	f_{\text{income tax}}(x)= \left\{
		\begin{array}{ll}
			0, & \text{if } x \leq \frac{18,200}{4}, \\
			0.16 \times \left(x - \frac{18,200}{4}\right), & \text{if } \frac{18,200}{4} < x \leq \frac{45,000}{4}, \\
			\frac{4,288}{4} + 0.30 \times \left(x - \frac{45,000}{4}\right), & \text{if } \frac{45,000}{4} < x \leq \frac{135,000}{4}, \\
			\frac{31,288}{4} + 0.37 \times \left(x - \frac{135,000}{4}\right), & \text{if } \frac{135,000}{4} < x \leq \frac{190,000}{4}, \\
			\frac{51,638}{4} + 0.45 \times \left(x - \frac{190,000}{4}\right), & \text{if } x > \frac{190,000}{4}.
		\end{array}\right.
		$$
		For identification purposes, $\tau_S(t)$ is set to $f_{\text{income tax}}(S(t))$, and $\tau_A(t)$ is obtained from equation \eqref{Eq3_2}. Note that income tax thresholds are not indexed in Australia.
		
		The tax rate on superannuation returns is 15\% before retirement and 0\% after retirement. Thus:
		$$
		\tau_F(t)= \left\{
			\begin{array}{ll}
				15\% r_F(t)F(t-1), & \text{if } t<T_r, \\
				0, & \text{if } t\geq T_r.
			\end{array}\right.
			$$
			Finally, the superannuation contribution tax rate is set at $15\%$, but the Low Income Super Tax Offset refund applies for yearly incomes below 37,000. Thus, the function $\tau_{\gamma}(t)$ is equal to 0 for $S(t)\leq \frac{37,000}{4}$, and to 15\% otherwise.
			
\subsection{More details on pension rules}\label{Appendix:pension}
During the accumulation phase, the only superannuation contributions considered are employer's mandatory contributions, fixed at rate $\gamma$. Voluntary contributions are excluded. The employer’s compulsory contribution, known as the superannuation guarantee, is $\gamma=12\%$, consistent with the recent update in July 2025.

During the decumulation phase, pension withdrawals $B(t)$ are at the household’s discretion. For simplicity, it is assumed that households withdraw the minimum required amount set by the government, given by:
$$B(t)=\left\{\begin{array}{ll}0,& \text{ for } t<T_r,\\b(t)F_{acc}(t), & \text{ for } t \geq T_r,\end{array}\right.$$
where the function $b(t)$ corresponds to the ATO's minimum drawdown rates, adjusted to quarterly frequency:
$$b(t)=\left\{\begin{array}{ll}\frac{5}{4}\%,& \text{ for } 168\leq t \leq 199,\\\frac{6}{4}\%,& \text{ for } 200\leq t \leq 219,\\\frac{7}{4}\%,& \text{ for } 220\leq t \leq 239,\\\frac{9}{4}\%,& \text{ for } 240\leq t \leq 259,\\\frac{11}{4}\%,& \text{ for } t \geq 260.\end{array}\right.$$

For households with insufficient assets and income, the government provides a pension income $G(t)$, known as the Age Pension. It includes three components: the basic age pension $G_{base}(t)$, pension supplements $G_{supp}(t)$, and rental assistance $G_{rent}(t)$ for non-homeowners. Eligibility and benefit amounts are determined by income and asset tests, with thresholds based on homeownership status and household composition (single vs. couple). Notably, a positive base age pension is required for pension supplements and rental assistance. Further details on the Australian age pension can be found in \cite{2013ChoSane} and \cite{2024Andreasson}.

The total guaranteed pension benefit is given by:
$$
G(t)=\left\{\begin{array}{ll}0,& \text{for } t<T_r,\\
	G_{base}(t) + G_{supp}(t) + G_{rent}(t),& \text{for } t\geq T_r, \text{ for non-home owners, provided } G_{base}(t)>0,\\
	G_{base}(t) + G_{supp}(t),& \text{for } t\geq T_r, \text{ for home owners, provided } G_{base}(t)>0,\\\end{array}\right.$$
The three components $G_{base}(t)$, $G_{supp}(t)$ and $G_{rent}(t)$ are the basic age pension, the supplements and rental assistance, respectively. They are given by:
\begin{eqnarray*}
	G_{base}(t) &=&6.5\times \min\{G^{\max}_{base},G_{asset}(t),G_{income}(t)\},\\
	G_{supp}(t) &=&6.5 \times \max \left\{G^{\min}_{S}(t),\frac{G_{base}(t)}{G^{\max}_{B}(t)}G^{\max}_{S}(t) \right\},\\
	G_{rent}(t)&=&6.5\times  \min\left\{\max\left\{0,\omega_{R}\times(R(t)-R^{\min}(t))\right\},R^{\max}(t)\right\},
\end{eqnarray*}
where the multiplication by 6.5 reflects quarterly payments, as all thresholds are given on a fortnightly basis. The functions $G_{asset}(t)$ and $G_{income}(t)$ are the asset and income tests determining the fortnightly pension reduction, and are defined as:
\begin{eqnarray*}
	G_{asset}(t)&=&\max\left\{0,G^{\max}_{B}(t) - \omega_{A}\times \left(A_{acc}(t) + F_{acc}(t) - W^{*}_{A}(t)\right)\right\},\\
	G_{income}(t)&=&\max\left\{0,G^{\max}_{B}(t) - \omega_{I}\times \left(I_{assessed}(t) - W^{*}_{I}(t)\right)\right\},
\end{eqnarray*}
where assessed income $I_{assessed}(t)$ is given by:
$$	I_{assessed}(t)=\zeta_1 \min\{A_{acc}(t) + F_{acc}(t),W^{*}_{D}(t)\} + \zeta_2 \max\{0,A_{acc}(t) + F_{acc}(t)-W^{*}_{D}(t)\}.$$
The functions $G^{\max}_{B}(t)$, $W^{*}_{A}(t)$, $W^{*}_{I}(t)$, $W^{*}_{D}(t)$, $G^{\min}_{S}(t)$, $G^{\max}_{S}(t)$, $R^{\min}(t)$ and $R^{\max}(t)$ are the thresholds that depend on household composition (single vs. couple), with $W^{*}_{A}(t)$ also depend on homeownership status. Their initial values, along with the rates $\omega_{asset}$, $\omega_{income}$, $\omega_{rent}$, $\zeta_1$ and $\zeta_2$, are provided in Table \ref{Table1}, and the time-$t$ values are the inflation-adjusted time-$0$ values.

\begin{table}[!h]\centering
	\begin{tabular}{lrr}\toprule
		Notation &  Singles & Couples\\\midrule
		$G^{\max}_{B}(0)$&  $1,051.3$&$1,585$\\
		$W^{*}_{A}(0)$ (non-homeowners)& $566,000$ &$722,000$\\
		$W^{*}_{A}(0)$ (homeowners)& $314,000$&$470,000$\\
		$W^{*}_{I}(0)$& $212$ &$372$\\
		$W^{*}_{D}(0)$& $62,600$ &$103,800$\\
		$G^{\min}_{S}(0)$&  $59.1$&$89$\\
		$G^{\max}_{S}(0)$&$97.7$&$147.2$\\
		$R^{\min}(0)$& $149.6$&$242.4$\\
		$R^{\max}(0)$&  $432.27$&$508.8$\\
		\midrule
		$\omega_{I}$&   $50\%$ & $25\%$ \\		$\omega_{A}$& \multicolumn{2}{c}{$0.3\%$} \\
		$\omega_{R}$&   \multicolumn{2}{c}{$75\%$} \\
		$\zeta_1$&   \multicolumn{2}{c}{$0.25\%$} \\
		$\zeta_2$&   \multicolumn{2}{c}{$2.25\%$} \\\bottomrule
	\end{tabular}
			\caption{\textbf{Thresholds and rates used to determine the guaranteed pension income $G(t)$.} \textit{ The top panel provides the thresholds, and the bottom panel provides the rates. All numerical values were obtained from the Australian Tax Office's website; see the main text for details. We note that all of the above values, in the top panel, are indexed by CPI.}}
	\label{Table1}
\end{table}

\end{document}